\documentclass[]{aa} % for the abstract without structuration (traditional abstract) 

\usepackage{graphicx}
\usepackage{txfonts}
\usepackage{graphicx}
\usepackage{natbib}
\usepackage{pifont}
\usepackage{booktabs}
\newcommand\bk{\bar\kappa}
\newcommand\gc{\gamma_{\rm c}}
\newcommand\gw{\gamma_{\rm w}}
\newcommand\gt{\gamma_{\rm g}}
\newcommand\Pc{P_{\rm c}}
\newcommand\Pg{P_{\rm g}}
\newcommand\Pw{P_{\rm w}}

\newcommand\vA{v_{\rm A}}
\newcommand\us{u_{\rm s}}

\newcommand\tacc{\tau_{\rm acc}}
\newcommand\pmax{p_{\rm max}}

\newcommand\Msb{M_{\rm SB}}
\newcommand\Vsb{V_{\rm SB}}
\newcommand\Esb{E_{\rm SB}}
\newcommand\Msol{\rm{M}_\odot}
\newcommand\dtsn{\Delta t_{\rm SN}}
\newcommand\Esn{E_{\rm SN}}
\begin{document}

\title{Time-dependent galactic winds}
\subtitle{II. Effects of boundary variations in the disc and galactic halo}

\author{E. A. Dorfi \inst{1} \and D. Steiner \inst{1} \and F. Ragossnig \inst{1}
   \and D. Breitschwerdt \inst{2} }
   
\offprints{E.A. Dorfi}

\institute{Universit\"at Wien, Institut f\"ur Astrophysik,
           T\"urkenschanzstr. 17, A-1180 Wien, Austria \\
           \email{ernst.dorfi@univie.ac.at}
           \and
           Zentrum f\"ur Astronomie und Astrophysik, Technische Universit\"at Berlin,
           Hardenbergstra{\ss}e 36, D-10623 Berlin, Germany \\
           \email{breitschwerdt@astro.physik.tu-berlin.de}  }

\date{Received date .......; accepted date .......}

\abstract
{Cosmic rays (CRs) are transported out of the galaxy by diffusion and advection due to streaming 
along magnetic field lines and resonant scattering off self-excited Magneto-Hydro-Dynamic (MHD) waves. Thus momentum is 
transferred to the plasma via the frozen-in waves as a mediator assisting the thermal pressure in 
driving a galactic wind.}
  {Galactic CRs (GCRs) are accelerated by shock waves generated in supernova 
remnants (SNRs), 
%%
%% DB: Ergänzung
%%
and they propagate from the disc into the halo. Therefore CR acceleration in the 
halo strongly depends on the inner disc boundary conditions. 
}
  {We performed hydrodynamical simulations of galactic winds in flux tube geometry appropriate 
  for disc galaxies, describing the CR diffusive-advective transport in a hydrodynamical 
   fashion (by taking appropriate moments of the Fokker-Planck equation) along with the 
   energy exchange with self-generated MHD waves.}
%%
%% DB: Die Plasmaphysik-Experten (Völk, Ptuskin, Zweibel etc.) behaupten, dass 
%nichtlineare (Landau-) Dämpfung die Wellenamplitude auf kleine Werte drückt und 
%die Energie der Wellen im Gas dissipiert wird (sog. Wellenheizung). Für die starke 
%Kopplung wäre eine kleine Wellenamplitude noch ausreichend, würde aber 
%energetisch keine Rolle mehr spielen. Darüberhinaus wäre man das Problem 
%los, dass bei delta B/B >> 1 die quasilineare Theorie nicht mehr gültig ist. 
%Für die Winde würde das bedeuten, dass sich wieder eher mehr Masse im Wind 
%befindet, da jetzt die Energie in Gasheizung geht, und das gamma = 5/3 ist, 
%statt 3/2 (für super-alfvenische Strömungen und 1/2 für subalfvenische) für die 
%Wellen. In meinem 91er Paper habe ich lineare Wellendämpfung getestet, was zu 
%nichtstationären Strömungen geführt hat. Ich weiß nicht wie umfangreich eure 
%Rechnungen sind, bzw. wie lange es dauern wird, solche Rechnungen durchzuführen. 
%Man würde die Wellengleichung und die Wellenterme  herausnehmen und stattdessen 
%den Wellenerzeugungsterm v_A grad P_c als Energiequelle in die Energiegleichung 
%übernehmen.Ich kann mir das mal genauer anschauen, wenn Du möchtest. 
%Ich habe daher sicherheitshalber einen kleinen Abschnitt eingefügt.
  {Our time-dependent CR hydrodynamic simulations confirm that the evolution of galactic winds with feedback 
  depends on the structure of the galactic halo. In case of a wind-structured halo, the wind breaks down after the last super nova (SN) has exploded.}
  {
%%
%% DB: Ergänzung 
%%  
  The mechanism described here offers a natural and elegant solution to
explain the power-law distribution of CRs between the `knee' and the 
`ankle'. 
The transition will be naturally smooth, because the Galactic CRs 
accelerated at SN shocks will be `post-accelerated' by shocks 
generated at the inner boundary and travelling through the halo. 
}

\keywords{Galaxies: evolution -- ISM: jets and outflows -- Galaxies: starburst
          -- supernova remnants -- cosmic rays}

\maketitle

\section{Introduction}

Galactic winds emerging on large scales from galaxies seem to be a ubiquitous
phenomenon in the Universe, as has been confirmed by numerous observations (for a recent review see \citet{ru:18}). 
%%
%% DB: Ergänzungen
%%
For example, it has been argued that at redshifts $z \sim 6 - 8$ 
starburst-driven galactic winds can drill channels in the gas 
distribution in order to allow photon escape and thus contribute to 
reionization \citep{sharma+17}. Substantial metal enrichment in low-redshift, circumgalactic halos as well as low-ionization metal absorption lines are also attributed to supernova (SN)-driven galactic wind bubbles \citep{LT:18}.
Moreover, numerical simulations suggest \citep[e.g.][]{Han:etal} that cosmic rays (CRs) can
drive large scale winds from high-redshift galaxies. In the past models of photon-driven winds have also been invoked, but as is well known the conversion of photon to gas momentum is low, typically less than 1\% (see \citet{DW:97}), 
and high dust opacities would be needed to drive a significant wind \citep{SO:15}.

Since in star-forming galaxies similar to the 
Milky Way there exists a rough energy equipartition between gas, CRs and magnetic fields, feedback processes in the interstellar medium (ISM) 
also depend on the dynamical action of CRs. To the extent that star 
formation or the growth of supermassive black holes in galaxies is limited by 
ISM feedback, CRs can play an active role in this respect \citep{ZW:17}.  

The realisation of the importance of CR feedback on galaxy evolution and star formation, as well as the back reaction on CR transport itself, has led to a wealth of recent studies (e.g. \citet{TP:19}, \citet{AB:18}, \citet{GN:18}, \citet{MO:18}). 
In this study we concentrate on regions above so-called superbubbles (SBs), generated by local sequential  
explosions, and we do not consider the large
scale outflows triggered by a global starburst  \citep[e.g.][]{Rus:2017}. 
The explosion energy of repeated SNe is shared between thermal heating, kinetic energy of the expanding bubble, and
acceleration of particles to relativistic energies by a first order Fermi mechanism 
\citep{Kry, Axf:etal, Bel:a, Bel:b, Bla:Ost:78}. The efficiency of CR acceleration in supernova remnants has been analysed with respect to the ambient magnetic field, which breaks the initial symmetry of the bubble and leads to a different acceleration efficiency for the resulting oblique shock (transition from parallel to perpendicular shock between equator and pole for an initially horizontal field).
It has been shown \citep{PP:18} that the overall efficiency is mainly due to the quasi-parallel part of the shock surface. In hydrodynamic simulations CRs are treated as a relativistic fluid with an adiabatic index $\gc=4/3$ and are coupled via the magnetic field and its fluctuations to the gaseous component. It has now been observed many times by radio synchrotron emission that CR electrons populate galactic halos, and their transport in a diffusion-advection halo can give useful insight into halo dynamics \citep{Bre:94, HK:18}. 

The galactic winds treated here are large scale outflows above a local concentration of SN explosions, which drive the flow by the combined action of thermal pressure, CR pressure, and
wave pressure of magnetic fluctuations. They provide a crucial mechanism to
transport metal-enriched material as well as magnetic fields and CRs into the intergalactic medium (IGM). 
Thus galactic winds are an important ingredient for the overall evolution of galaxies and their associated 
ISM, including star formation,  due to mass and energy losses. It has been shown that magnetized galactic winds also 
foster angular momentum loss \citep{Zirak:96}, which can amount to 40\% over a lifetime of 
ten billion years in the case of the Milky Way.

To be more specific, we want to study how locally concentrated outflows are generated. After break-out of the expanding bubble into the galactic halo, the remaining SNe
exploding sequentially within the bubble release energy until the reservoir of SN progenitors
is exhausted, typically when all stars above $8 \, \Msol$ have exploded (a lower limit of $6 \, \Msol$ is used when taking into account the possibility of electron-capture SNe). Hence, we get a time-dependent boundary between the wind region and the bubble underneath. 
%%DB: geändert, sonst könnte man stationäre Lösungen ganz vergessen
Such galactic winds are traversed by shock waves and, strictly speaking, the changing
inner boundary conditions will allow stationary outflows only in a time-averaged sense.
The typical lifetimes of OB-associations of several $10^6 - 10^7\,$ years \citep{KM:87} are short compared
to the flow times through a galactic wind, and stationary winds require power from several OB-associations due to propagating star formation in the disc. 
This is in contrast to starbursts with timescales of several $10^8\,$ years where 
winds can be mainly driven by the thermal pressure. In contrast, in the case of SB-driven galactic winds,
CRs play a vital role in driving the outflow and we also 
examine the possibility that particle acceleration in shocks propagating down the density gradient in galactic wind flows can boost the particles even to higher energies, in particular if the  
main site of acceleration is not too far from the
galactic disc, as emphasized in \citet{DB1}. Since downstream is the region facing the galactic disc, it is easy for accelerated particles to be convected away from the shock, diffusing towards the disc.
The coalescence of repeated shock waves 
generated by sequential SNe provides a natural environment for such successive re-acceleration events. 
In this paper we can now closely link the SB evolution to the 
time-dependent wind structure and explore in detail the re-acceleration site of
energetic CRs that are typically less than $15\,{\rm kpc}$ above the galactic plane.

%%DB: hier wäre das 91er-Paper das bessere Zitat
For further details on the physical equations, assumptions and numerical 
method we refer to \citet{DB91} or \citet{DB1}. 
The main improvement compared to our earlier simulations is due to
the time-dependent boundary conditions that closely link the galactic disc 
parameter to the flux tube properties as outlined in detail in Sect.~\ref{s.model}.

A different model for CR acceleration between the 'knee' and the 'ankle' has been proposed by \citet{VZ:04}, 
who use the propagation of compressional waves through the halo generated by the spiral 
density shock waves in the disc. The pattern speed is different from the Galactic rotation both in the disc and the halo. 
Therefore magnetic field lines rooted in the dense, low, plasma-beta regions in the disc will slip through weak CR-modified shocks, arising from the collision of Galactic wind flows from regions with higher velocity with those of low velocity gas flows. These so-called slipping interaction regions are assumed to be at a vertical distance of about 20 kpc from the Galactic midplane and should be responsible for the re-acceleration of Galactic CRs to energies up to $10^{18} \, {\rm eV}$. While back diffusion of accelerated particles will be much easier than from inwards-facing-galactic-wind termination shock (downstream is away from the disc) at distances more than 100 kpc, we note that in our model \citep{DB1} the re-acceleration region is much closer (below 10 kpc) to the disc, with the shocks' downstream regions pointing towards the Galactic disc.

There are also models that favour acceleration in supernova remnant (SNR) shocks by the first order Fermi mechanism in the disc, arguing for a higher magnetic field strength due to SN explosion into a stellar-wind-compressed ISM \citep{VB:88} or due to wave growth upstream of the shock as a result of CR streaming \citep{BL:01}. The spectral break beyond the knee has been assigned in the latter case to poor source statistics. 

Here we focus on the origin and continuity of the galactic wind flow emanating from the SN and SB sources in the disc, specifically on the time variations of the inner Galactic wind boundary.
The paper is structured as follows. In Sect.~\ref{s.model} we
describe the physical properties of our model, in particular the
evolution of the inner boundary during the galactic wind outflow. 
In Sect.~\ref{s.res} our results from time-dependent
galactic wind simulations for the Milky Way are presented to show 
how the shock waves generated by the repeated SN explosions coalesce to
a single strong shock propagating into the intergalactic space. In particular 
we emphasize CRs which are accelerated bby these successive shock waves.
A discussion and our conclusions close the paper in
Sect.~\ref{s.con}.

\section{Boundaries of galactic wind models}
\label{s.model}

\subsection{Kompaneets approximation}
\label{s.komp}

\begin{figure}   %                                                         Fig. 1
%  \resizebox{\hsize}{!}{\includegraphics{Plots_Paper_v2/ini_mod_3.eps}}
 \resizebox{\hsize}{!}{\includegraphics{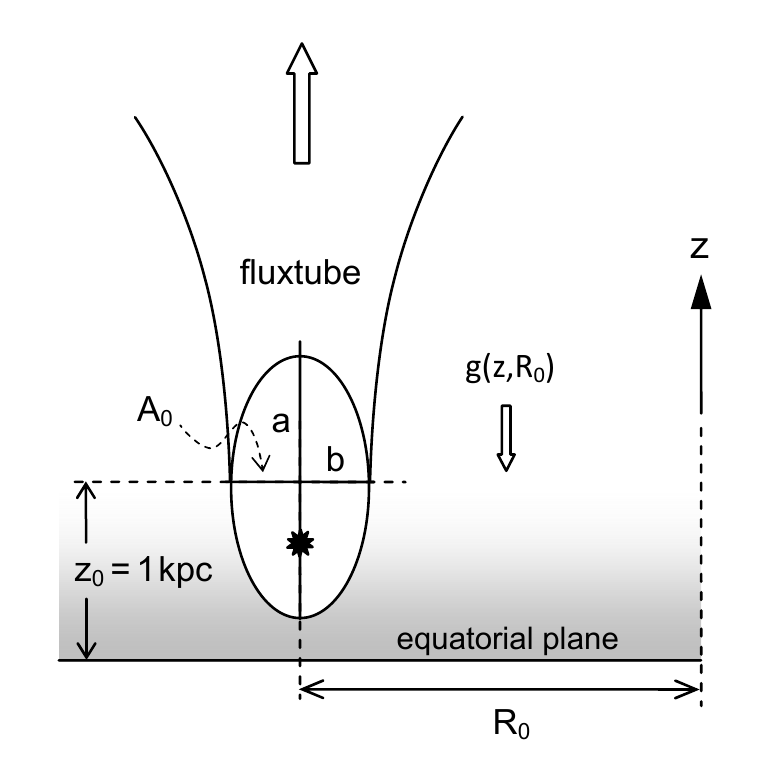}}
  \caption{Sketch of the initial configuration adopted for the 
           galactic wind models. The SB is approximated by a
           Kompaneets solution until break-out, where $z_0 = 1\,{\rm kpc}$ fixes
           the location of the inner boundary conditions of the flux tube. All
           physical variables depend on time $t$ and on the distance from the
           Galactic plane $z$. The gravitational acceleration $g(z,R_0)$ also depends
            on $R_0$, the Galactocentric distance of the flux tube (see text for
           more details).}
  \label{f.ini_mod}
\end{figure}

As described in \citet{DB1}, the one-dimensional time-dependent
treatment of galactic outflows in flux tube geometry
requires initial as well as boundary conditions. In this paper
we  use a simplified description of the evolution of the SB to 
generate the initial conditions at the bottom of the flux tube, located at
$z_0 = 1\,{\rm kpc}$. According to \citet{Kompa} we calculate an 
explosion within an exponentially stratified ISM using
\begin{equation} 
  \rho(z) = \rho_{00}e^{-z/h_z} \:,
\end{equation} 
where $\rho_{00}=5\cdot 10^{-24}{\rm g\,cm^{-3}}$ specifies the gas density 
at the equatorial plane and 
$h_z=80\,{\rm pc}$
denotes the vertical scale height. We can link the
sequence of SN explosions to the evolution of a SB until break out,
and adopt the SB properties as initial conditions at the
bottom of the galactic wind driven by the gradients of thermal pressure, CR
pressure, and Alfv\'enic wave pressure \citep[e.g.][]{DB91}. The overall implementation
of the SB evolution is sketched in Fig.~\ref{f.ini_mod}. As suggested
in the literature \citep{Mac, BB:13}, the volume surrounded by the expanding shock wave 
is almost indistinguishable from an ellipsoid 
and can therefore be approximated by
\begin{equation}
\Vsb = \frac{4\pi}{3}a^2(t)b(t) \:,
\end{equation}
where $a$ and $b$ are the semi-major and semi-minor axes of the rotationally symmetric ellipsoid. At
breakout time $t_0$ we kept $a(t_0)$ and $b(t_0)$ fixed and
the lateral extension of the super bubble at this time determines
the area of the flux tube at the lower boundary through 
\begin{equation}
A_0 = \pi b(t_0)^2  \:.
\end{equation}
The cross section $A(z)$ of the flux tube is given by
\begin{equation}           \label{e.az}
A(z) = A_0\left[ 1 + \left( \frac{z}{z_b} \right)^2 \right]  \:,
\end{equation}
where we adopt $z_b=15\,{\rm kpc}$ as typical scale length for widening the
flux tube towards a more spherical geometry. All computations are performed
for a Galactocentric distance of $R_0 = 10\,{\rm kpc}$.

%  EAD: Teil auskommentiert
% The parameters of the Kompaneets model are determined by the explosions of $25$ Supernovae within a
%  time of $12\cdot 10^6\,{\rm years}$ leading to a typical explosion interval of
% $\dtsn= 4.8\cdot 10^5\,{\rm yr}$. We adopt these values
% as averages for the size and life time of OB-associations. 
%%DB: ich würde den Teil darüber weglassen und generell die Abschätzung unten benutzen.

This can be estimated as follows. According to \cite{FB:06}, the main sequence lifetime of a star can be calculated by $\tau_{ms} = \tau_0 (m/M_\odot)^{-\beta}$, 
with $\tau_0 = 1.6 \, 10^8$ yr and $\beta = 0.932$. If we take $m_l = 
8\, \Msol$ to be the lowest and $m_u = 60\, \Msol$ the highest SN progenitor 
mass, then the explosion interval is the range of explosion times divided 
by the number of SN progenitors between $m_l$ and $m_u$, $N_\ast$, that is,  
$\Delta\tau = \tau_0 M_\odot^{-\beta} (m_l^{-\beta} - m_u^{-\beta})/N_\ast$, 
where $N_\ast$ depends on the richness of the cluster. Thus $\Delta\tau \approx 
1.95 10^7\, {\rm yr}/N_\ast = 4.8 10^5$ yr for $N_\ast = 40$. A more detailed 
calculation taking into account the initial mass function can be found in 
\citet{BF:16}.

\begin{figure}   %                                                         Fig. 2
%
% \resizebox{\hsize}{!}{\includegraphics{Plots_Paper_v2/plot_kompaneets_paper_energy_amajor.eps}}
  \resizebox{\hsize}{!}{\includegraphics{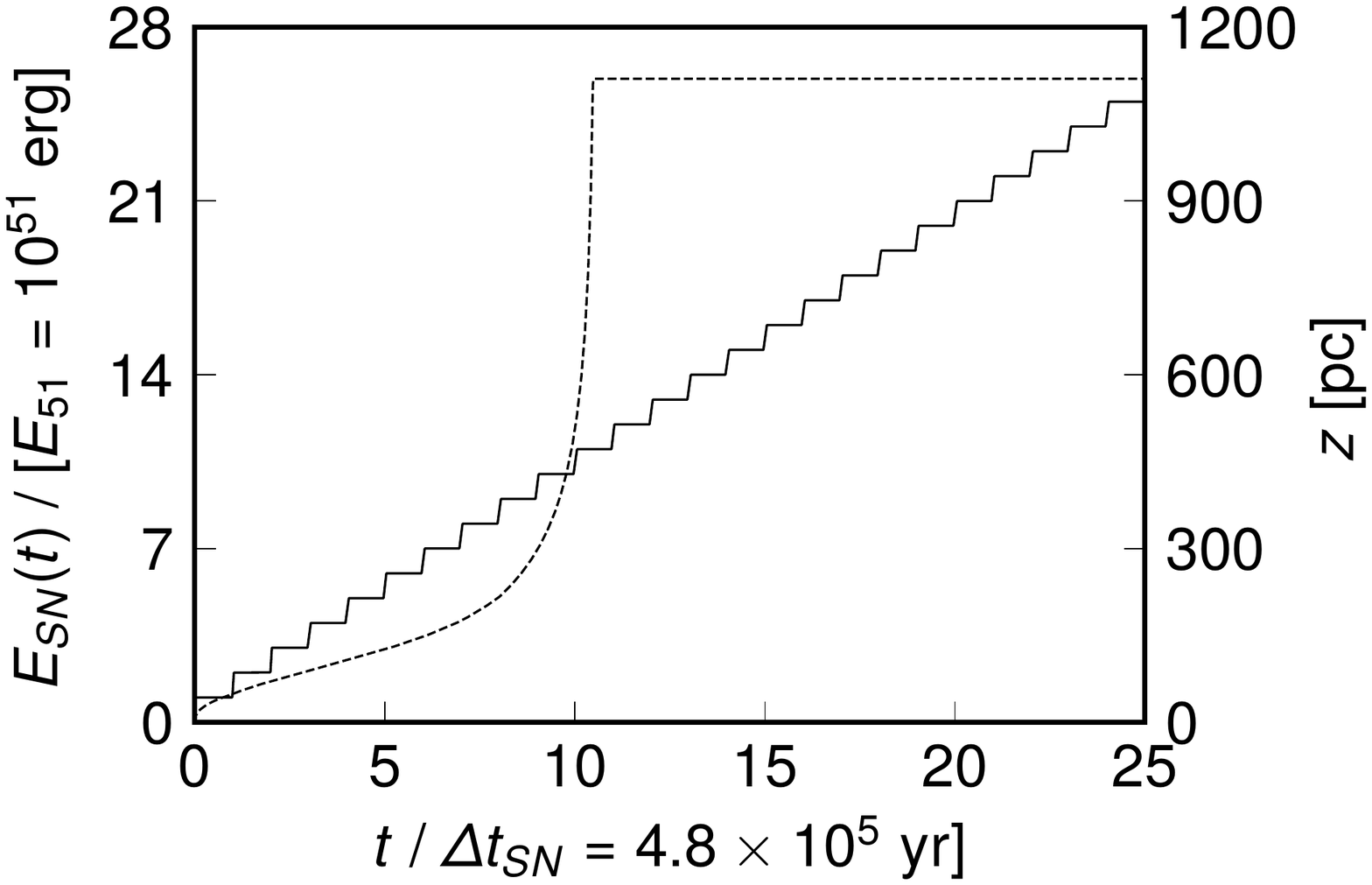}}
  \caption{Total energy increase of the repeated SN explosion (solid line) in units of
           $\Esn=10^{51}{\rm erg}$ as a function of time in units 
           $\dtsn = 4.8\cdot 10^5{\rm yr}$, the time between the individual SN explosions. 
           The temporal expansion of the semi-major axis $a(t)$ based on the 
           Kompaneets ellipsoid in units of the explosion interval $\dtsn$ clearly shows
           the time $t_0=10.3\dtsn$ when the base of the wind expansion 
           is reached at $z_0=1\,{\rm kpc.}$}.        
  \label{f.ini_komp}
\end{figure}

%(ZITATE, Begr\"undung oder Diskussion f\"ur diese Anfangsbedingungen).
Each SN explosion releases $\Esn=10^{51}\,{\rm erg,}$ 
which results in a step-wise increase of the total energy $\Esb(t)$ of the SB (see
Fig.~\ref{f.ini_komp}, full line). The thermal energy of the bubble is given
by $90\%$ of the total explosion energy and the remaining $10\%$ is used for 
acceleration of CRs within the SB, meaning~$E_{\rm CR}(t) = 0.1\,\Esb(t)$ 

From Fig.~\ref{f.ini_komp} we can infer the evolution of the 
semi-major axis $a(t)$ (dotted line) of the ellipsoid describing the SB.
We see that after $t_0=10.3\,\dtsn = 4.94\cdot 10^{6}\,{\rm yr}$ 
the bubble opens rapidly into a galactic wind and we obtain $b(t_0)= 251\,{\rm pc}$ and 
$A_0 = \frac{1}{2} \pi b^2(t_0) = 9.887\cdot 10^5{\rm pc^2}$ as starting values 
for the subsequent time-dependent wind computations 
(the factor $1/2$ arises due to the assumption of a non-constant velocity cross 
section through $A_0$). 
The first $13$ out of $25$ SN explosions are needed to develop 
this inner structure necessary to initiate the outflow.

%%DB: ich habe immer semi-major axis geschrieben statt major half-axis
%% Es wird wohl kein Zufall sein, dass break-out gerade dann passiert, wenn z= 1 kpc ist. Habt ihr die Parameter entsprechend gewählt?
%%

All general features of these galactic wind models have been described in detail by
\citet{DB1} and without repeating their properties we note that the galactic 
potential defining $g(z,R_0)$ has been calculated according to \citet{MN}. The physical
equations are given by the equation of continuity and the equation of motions containing
the pressure gradients of thermal gas, CRs, and Alfv\'enic waves, and the
system is closed by three energy equations for the gas, the CRs, and the waves. The
CRs are treated as a relativistic fluid with $\gc=4/3$ and include a
diffusive term with a mean diffusion coefficient $\bk$ describing the acceleration of
particles by a first-order Fermi mechanism \citep[e.g.][]{Dru}. 
%%
%% DB: Flussröhrengeometrie ist NICHT orthogonal; ich habe den metrischen Tensor 
%% mal abgeleitet, er hat Nebendiagonalelemente
%% --> Orthogonal wurde gestrichen
%%
All equations are
written in flux tube geometry and depend on the distance from the galactic 
plane $z$ and time $t$.

As mentioned before, the temporal evolution of galactic winds depends not only on 
the variations at the inner boundary, but also on the
initial vertical structure of the medium. Therefore we explore basically two
scenarios where the initial background structure corresponds either to a
stationary wind ({\sf Model W}) or the wind is evolving into an
almost hydrostatic vertical stratification ({\sf Model H}). Table~\ref{t.ini_mod}
summarizes the properties of both initial structures.

\begin{table}
\caption{Values at the lower boundary of the two initial background models. The time-dependent
         galactic winds evolve into these background models.}
\label{t.ini_mod}
\centering
\begin{tabular}{l|c|c}
 \toprule
                                       &  {\sf Model W}           & {\sf Model H} \\
 \midrule
 $P_{{\rm g},0}\,[{\rm dyn\, cm^{-2}}]$ & $2.76\!\cdot\!10^{-13}$ & $2.76\!\cdot\!10^{-14}$ \\
 $P_{{\rm c},0}\,[{\rm dyn\, cm^{-2}}]$ & $1.00\!\cdot\!10^{-13}$ &  $1.00\!\cdot\!10^{-14}$ \\
 $P_{{\rm w},0}\,[{\rm dyn\, cm^{-2}}]$ & $3.98\!\cdot\!10^{-16}$ &  $4.00\!\cdot\!10^{-17}$ \\
 $\rho_0\,[{\rm g\,cm^{-3}}]$           & $1.67\!\cdot\!10^{-27}$ &  $1.67\!\cdot\!10^{-27}$ \\
 $u_0\, [{\rm km/s}]$                   & $9.88$                    &  $0.11$  \\
 $\dot M_0\, [{\rm g\, s}^{-1}]$          &  $1.55\!\cdot\!10^{22}$       & $1.68\!\cdot\!10^{20}$ \\
 \bottomrule
\end{tabular}
\end{table}

The initial background {\sf Model W} is similar to the stationary reference model 
%%DB: ich denke, Du meinst das 91er Paper
of \citet{DB91} and the
time-dependent evolution on timescales larger than $10^7\,$years is similar 
to the simpler models already presented \citep{DB1}. In those earlier models the pressures at
the inner boundary have been increased in one single step by a factor of $10$  to
study the propagation of a strong shock wave into the galactic wind.  

For the initial background {\sf Model H} we have reduced 
all pressures by a factor of $10$ and hence the
outflow velocity is reduced by a factor of $90$ compared to {\sf Model W}. The velocity at the
inner boundary in now only $u_0=0.11\,{\rm km/s}$ and $u_0 \ll \vA (= 69\,{\rm km/s})$.
Since the density is kept constant, 
corresponding to $n_0=1\,{\rm cm^{-3}}$, the mass loss is reduced by the same
factor of $90$ and is therefore negligible. The density of this model drops by
almost six orders of magnitude within the inner $50\,{\rm kpc}$. 

Because by definition all wind models must have non-zero initial velocity, a high-speed tenuous wind with $\rho \simeq 10^{-34}{\rm g\,cm^{-3}}$  and a velocity on the order of $1000\,{\rm km/s}$ at distances of $300\,{\rm kpc}$ will always be the result, but with negligible mass loss.
%Since we do not
%start at the inner boundary with $u_0=0$, an extremely tenuous wind 
%of $\rho \simeq 10^{-34}{\rm g\,cm^{-3}}$ can reach a
%velocity of about $1000\,{\rm km/s}$ at distances of $300\,{\rm kpc}$.
However, as seen for example~in Figs.~\ref{f.ini_val_mod_global} and \ref{f.rad_loss_W}, 
the time-dependent
solutions are not influenced by the initial models at such large distances 
and we can concentrate on 
flow features close to the galactic plane. For {\sf Model H} we expect a faster 
evolution due to
larger pressure gradients after the bubble outbreak. The lower gas density in the
outer regions will also accelerate the wind evolution because less 
material close to the galactic disc has to be pushed
away by the emerging wind. As seen in Eq.~\ref{e.flux}, even the case of $u_0=0$ can lead to
CR losses owing either to streaming with the Alfv\'en velocity $\vA$ or to a
diffusive flux with $\bk$.

The outer boundary is located at $r=300\,{\rm kpc}$ where a simple outflow boundary condition 
is applied,~$\partial u/\partial z = 0$, and all other physical variables are 
advected outwards. Numerical tests have shown that the wind solutions are not affected when
moving the outer boundary from $1000\,{\rm kpc}$ \cite[]{DB1} to the current value 
of $r=300\,{\rm kpc}$. Adopting such a spatially smaller computational domain allows 
a better resolution of the various flow features.
%%
%% DB: and faster computation?
%%

\subsection{Computation of inner boundary values}
\label{s.num}

The total mass within the SB $\Msb$ is given through
\begin{equation} \label{e.msb}
  \Msb(t) = M_{\rm ini} + \int_0^t \dot M_{\rm SN}\,dt - \int_0^t\dot M_{\rm GW}\,dt
,\end{equation}
where $\dot M_{\rm SN}$ denotes the mass input from stellar winds and the
repeated SN explosions, and $\dot M_{\rm GW}= A_0\rho_0 u_0$ denotes the mass loss
by the galactic wind. However, since
the mass encompassed in the SB shell is on the order of $3 \cdot 10^4 \Msol$, it will dominate  the ejecta mass by a large factor.  
% EAD: 
%  Version von DB verwendet
%
% the initial mass of the SB
% corresponds to $M_{\rm ini} = 3.xx\cdot 10^4\,\Msol$ and we
% can neglect the gas contribution from the SN-explosions since 
% $M_{\rm ini} \gg 25*M_{\rm SN}$.
%%
%% DB: was ist hier genau gemeint? Die SB-Masse ist tatsächlich nur die Anzahl der 
% SN-Vorläufersterne (- 2 Msol %für Neutronensternmasse pro Stern). Mit den 3 10^4 Msol ist das in der äußeren 
%Schale aufgeschobene ISM gemeint? Vielleicht sollte man dann besser schreiben: 
% The mass encompassed in the SB shell is of the order of 3 10^4 Msol and therefore dominates the ejecta mass by a large factor.  
% Oder ist hier die Masse aus heißem Gas aus den GW-Anfangsbedingungen gemeint, das 
%letztlich aus der Heizung früherer Generationen von SNe herrührt? Dann sollte man schreiben: 
% The SB mass originates from heating of the SB volume by previous SN generations and encompasses a 
% mass of 3 10^4 Msol, dominating the ejecta mass by a large factor.
%%
After the initial vertical opening of the SB, 
we set the area $A_0$ of the flux tube to remain constant. Equation~(\ref{e.msb}) also contains the feedback from the galactic wind transporting gas into the galactic halo.
The gas density $\rho_0(t)$ at the inner boundary of the flux tube can 
be calculated from the mass balance (Eq. \ref{e.msb}) and volume $\Vsb$ of the SB,
\begin{equation} \label{e.rho0}
  \rho_0(t) = \frac{\Msb(t)}{\Vsb}
.\end{equation}

%%
%% DB: müsste nicht wenigstens bei t=0 der CR-Druck 0.1 mal Gasdruck sein?
%% die beiden Bilder in Fig. 3 oben links und rechts geben das nicht wieder;
%% hängt das evtl. damit zusammen, dass bei t=0 schon 8 SNe explodiert sind? Vielleicht sollte man das in einem kurzen Satz erwähnen
%%
%% In Model W geht die Massenverlustrate (und damit auch u0) ab t = 6 10^6 wieder zurück; das liegt sicher daran, dass keine SN mehr explodiert
%% aber sollte das nicht mit einer Zeitverzögerung passieren, bis die SB evakuiert ist? Oder anders ausgedrückt: wie sinnvoll ist es alle Variablen 
%% bis auf u0 konstant zu halten? Ist es schwierig die Variablen,  z.B. durch Kühlung und Evakuierung des Gases abnehmen zu lassen?
%%
\begin{figure}   %                                                         Fig. 3
%
%  \resizebox{\hsize}{!}{\includegraphics{Plots_Paper_v2/plot_IBC_NOML_paper_K28.eps}}
 \resizebox{\hsize}{!}{\includegraphics{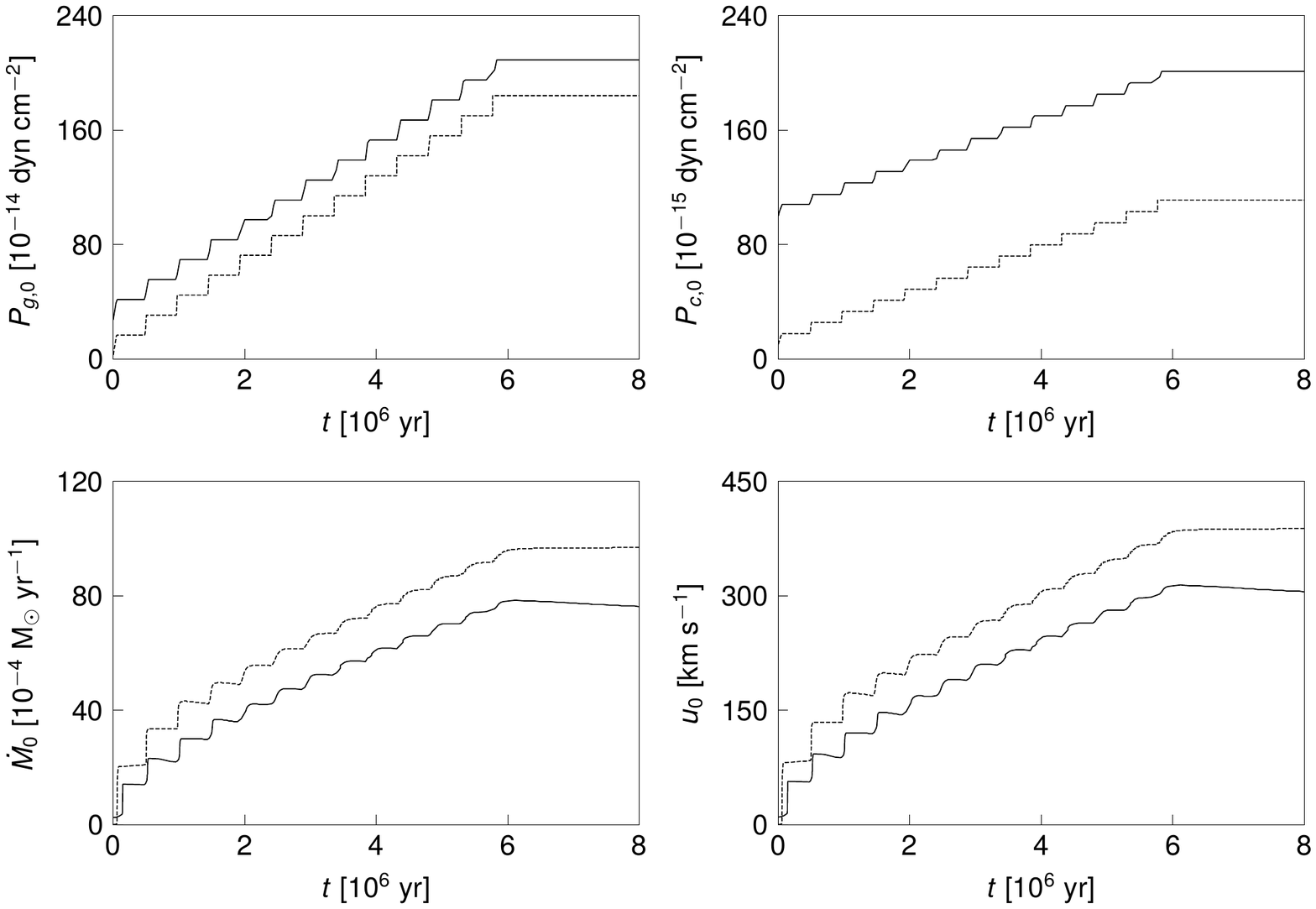}}
  \caption{Gas pressure $P_{{\rm g},0}$ at the inner boundary ($z_0=1\,{\rm kpc}$)
           of the flux tube for the wind background model {\sf Model W} (full line) and the almost 
           hydrostatic background {\sf Model H} (dashed line) as
           a function of time. After $6\!\cdot\! 10^6{\rm yr}$ the last SN exploded within
           the SB and we keep the values of the boundary values fixed. Every
           $\dtsn = 4.8\cdot 10^5{\rm yr}$ and amount of 
           $0.9\,\Esn=0.9 \cdot 10^{51}{\rm erg}$ is added to increase $P_{{\rm g},0}$.
           The CR pressure $P_{{\rm c},0}$ increases by the usage of $0.1\,\Esn$
           at every SN explosion. The lower panels exhibit the mass loss by 
           the galactic wind, that is,~$\dot M_0 = A_0\rho_0 u_0$ and the velocity at $u_0$.
           We note that during the phase of repeating SN explosions, {\sf Model H} generates 
           higher mass loss rates because the winds
           expand into a more tenuous medium than {\sf Model W}.}      
  \label{f.ini_val_mod_no}
\end{figure}

In analogy, we can calculate the thermal pressure at the boundary of the SB using 
\begin{equation} \label{e.ene_sb}
  P_{{\rm g},0}(t) = \frac{\gt -1}{\Vsb}
          \left[ E_{\rm SB,ini} + \int_0^t \dot E_{\rm SN,th}\,dt - \int_0^t\dot E_{\rm GW}\,dt \right] \:,
\end{equation}
where we have assumed $\gt=5/3$ and $E_{\rm SN,th} = 0.9\,\Esn = 0.9 \cdot 10^{51}{\rm erg}$
throughout all computations. The last term is given by integration of the outflow of 
hot gas through the 
surface $A_0$ with velocity $u_0$. 
%%
%% DB: E_therm taucht nicht in der Gl. auf, dafür E_SB,ini und E_GW; ich denke, man sollte sie an dieser Stelle definieren; was ist dE_GW/dt:
%% kinetische oder thermische Energierate?
%%

A similar Eq. (\ref{e.ene_sb}) holds for the CR pressure
at the inner boundary of the flux tube but the loss term has to be modified 
according to 
\begin{equation} \label{e.flux}
 \dot E_{\rm c,GW} = A_0 F_{{\rm c},0} = A_0\left[ \frac{\gc}{\gc-1}({u}_0+\vA)P_{{\rm c},0} -
     \left.\frac{\bk}{\gc-1}\frac{\partial\Pc}{\partial z}\right\vert_0 \right] 
,\end{equation}
which includes the streaming of CRs with the combined velocity of 
the gas flow together with the Alfv\'en speed $u_0+\vA$ and the diffusive losses
determined by the mean diffusion coefficient $\bk$. Assuming a magnetic field
of $B_0=10^{-6}\,{\rm G,}$ we obtain an Alfv\'en velocity of $\vA= 69\,{\rm km/s.}$ 
We assume that $10\%$ of
the SN explosion energy is converted into CRs, therefore we add 
$E_{\rm c} = 0.1\,\Esn$ at every time interval of $\dtsn = 4.8\cdot 10^5{\rm yr}$.

The Alfv\'enic wave pressure $P_{{\rm w},0}$ plays, for the initial conditions, only a minor
role (compare with Table~\ref{t.ini_mod}). 
%since we assume a fluctuating component of the field of $\delta B = 0.1\,B_0$
%and calculate $\Pw = \delta B^2 / 4\pi = ...$ . 
The temporal evolution at the inner boundary
is given by an equation like Eq. (\ref{e.ene_sb}) and in the case of a  
wind the inner boundary value is reduced due to   
\begin{equation} \label{e.wave_loss}
 \dot E_{\rm w,GW} = A_0\left[\frac{\gw}{\gw-1}({u}_0+\vA)P_{{\rm W},0}\right] \:.
\end{equation}

The temporal evolution of all loss terms like $\dot M_{\rm GW}$ and $\dot E_{\rm GW}$ , as well
as the different contributions of advection and diffusion in Eq.~(\ref{e.flux}), are
presented in Fig.~\ref{f.int_loss}.
Again, for more details on the physical system of equation we refer to \citet{DB91} or
\citet{DB1}. 

Figure~\ref{f.ini_val_mod_no} exhibits the
temporal change of the gas pressure $\Pg(t)$, the
CR pressure $\Pc(t)$, the mass loss rate $\dot M_0(t),$ and the
gas velocity $u_0(t)$ at the inner boundary in the time interval up to $8\!\cdot\!10^6{\,\rm years}$.
We set $t=0$ at the time when the first eight SNe have opened 
the SB. s
All physical variables increase in time and after the last SN 
added the energy to the SB at $t> 6\cdot 10^6\,$years, we kept all variables fixed
except the velocity $u_0$ , which has to adapt to the wind solution.
In the velocity panel (lower right) we can see how the
flow reacts to the change from increasing pressures $\partial P/\partial t>0$ to a
constant pressure by a reduction of the velocity $u_0(t)$ after $6\cdot 10^6\,$years.
The change of the inner boundary to a constant value generates a rarefaction wave that
travels at sound speed in the downstream region of the shocks. Such a non-linear wave
reaches the shock wave from behind and reduces its strength \citep{CF} therefore coupling
the wind solution to the velocity value $u_0(t)$ at the inner boundary. 
Since $\rho_0$ is kept constant, also the mass loss $\dot M_0 = A_0\rho_0 u_0(t)$ is
influenced by the velocity decrease.
This effect also becomes visible on larger temporal scales as plotted in 
Fig.~\ref{f.ini_val_mod_no_global}.

The acceleration of CRs by a first-order Fermi process is included 
by adopting a mean diffusion coefficient 
of $\bk = 10^{28}\,{\rm cm^2\,s^{-1}}$ , which mediates the jumps of 
$P_{{\rm c},0}$ at the inner boundary (Fig.~\ref{f.ini_val_mod_no}, upper right panel). 
We note that the particle pressure 
is typically a factor of $10$ lower than the gas pressure and we have adopted
the different scaling factors for the plots. However, as the flow
expands into the intergalactic medium the adiabatic index of
the energetic particles of $\gc=4/3$ is responsible for smaller adiabatic losses 
compared to the thermal gas ($\gt=5/3$). The adiabatic index, together with the diffusion 
of particles that led to mediating gradients, both led to an important contribution on larger scales,
as previously described in \citet{DB93} or \citet{DB1}. 

% visible in Figs.~\ref{f.rad_loss_A}, \ref{f.rad_loss_B}.

Figure \ref{f.ini_val_mod_no_global} plots the temporal evolution of the
boundary conditions over a time interval of $10^8\,$years, much longer than 
the SB evolution of  $6\cdot 10^6\,${years. The lower 
extragalactic density ({\sf Model H}, dotted line) leads to faster outflows 
with higher mass loss rates for the
first $6.4\cdot 10^7\,$years. A maximum of the mass loss rate 
$\dot M_{\rm max} = 1.03 \cdot 10^{-2}\Msol\,{\rm yr^{-1}}$ is reached
at $3.68 \cdot 10^7\,${years.

\begin{figure}   %                                                         Fig. 4
%
% inner boundary with no back reaction, global
%
%  \resizebox{\hsize}{!}{\includegraphics{Plots_Paper_v2/plot_IBC_NOML_paper_K28_wide.eps}}
  \resizebox{\hsize}{!}{\includegraphics{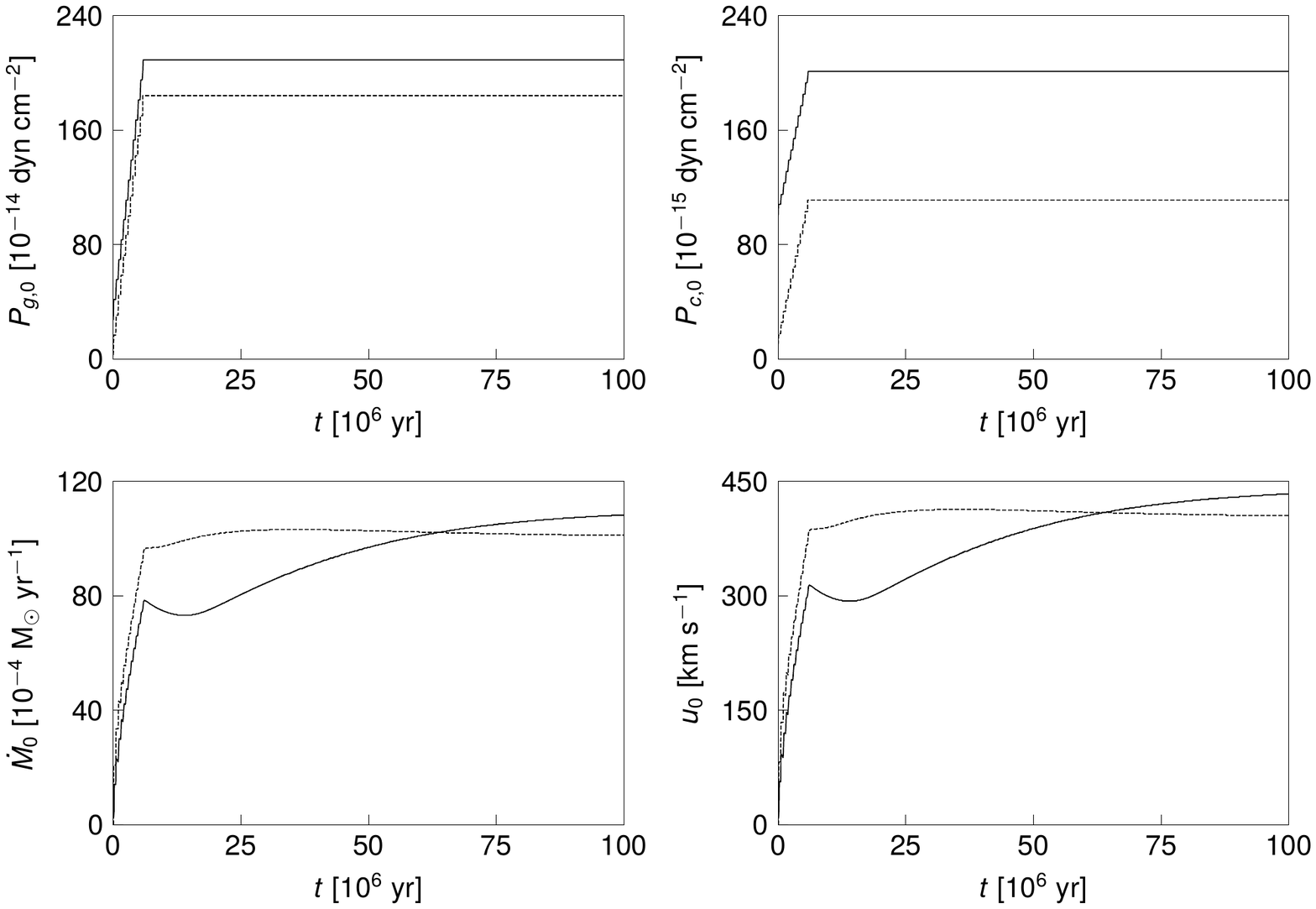}}
  \caption{Gas pressure $P_{{\rm g},0}$ at the inner boundary ($z_0=1\,{\rm kpc}$)
           of the flux tube for {\sf Model W} (full line) and {\sf Model H} (dashed line) as
           a function of time over $10^8\,$years. After $6\!\cdot\! 10^6{\rm yr}$ the last SN exploded and we keep the values of 
           the boundary values fixed, except the velocity $u_0$. The lower panel (right)
           illustrates how the velocity reacts to the change of the inner boundary from
           increasing pressures to constant pressures after $6\!\cdot\! 10^6\,$years.
           The lower panels exhibit the mass loss by
           the galactic wind, that is,~$\dot M_0 = A_0\rho_0 u_0$ and the velocity at $u_0$.
           We note that {\sf Model H} (dotted line) generates initially higher mass 
           loss rates because the wind
           expands into a more tenuous medium than {\sf Model W} (see text for more
           details).}      
  \label{f.ini_val_mod_no_global}
\end{figure}

As the temporal evolution goes on, the initially larger values of
the thermal pressure $P_{{\rm g},0}$ and the CR pressure $P_{{\rm c},0}$ of
{\sf Model W} steadily increase the outflow velocity at the inner boundary $u_0$, 
which results in
a further acceleration of the flow at later times. The stationary outflow at 
times longer than $10^8\,$years is about $20\%$ larger for {\sf Model W} (full line) 
than {\sf Model H} (dotted line). Since the outer boundary is located at $300\,{\rm kpc}$ the
flow time up to this distance is $3.83\cdot 10^8\,$years and $1.03\cdot 10^8\,$years,
respectively, which is the minimum needed to establish stationary
solutions. The evolution on these large timescales
characterized by a common shock wave is similar to the simple models of \citet{DB1} where
from the beginning of a single shock has been initiated.

Comparing the upper panels of Fig.~\ref{f.ini_val_mod_no_global},
with constant pressures after the last SN explosion ($t>6\cdot 10^6\,{\rm yr}$), to lower
panels (mass loss rate $\dot M_0$ and inner outflow velocity $u_0$), we can deduce the
timescale by which a stationary solution is reached. \sf Model W} (full line) 
is characterized by lower
velocities and therefore the time to propagate the constancy of the inner boundary values takes
about $10^8\,$years, meaning that~$10^8\,$years are necessary to establish a stationary solution
within our computational domain of $300\,{\rm kpc}$. As mentioned before, the decrease of the
velocity $u_0$ (and mass loss rate $\dot M_0$)
around $18\cdot 10^6\,{\rm yr}$ is caused by a rarefaction wave induced 
through the change of the inner boundary values. This feature will become even more
pronounced in the cases where the feedback between the SB and the wind losses
are taken into account (see Sect.~\ref{s.res} and in particular Fig.~\ref{f.rad_loss_W}). 
{\sf Model W} (dotted line)
develops faster and generates less massive winds within large density
gradients. These gradients also mediate the rarefaction wave and 
the change of the inner boundary conditions is less important.
Already after about $25\cdot 10^6\,{\rm yr}$ we find almost no
further changes in the global outflow structure.

Clearly, the concept of keeping the inner boundary 
conditions constant after the last SN exploded or neglecting
the loss terms is rather artificial, but it is useful to
separate and illustrate physical effects occurring in time-dependent galactic outflows.  
As seen in Sect.~\ref{s.long} with corresponding 
Figs.~\ref{f.ini_val_mod}, \ref{f.ini_val_mod_global}, and \ref{f.int_loss}, the feedback of the 
mass and energy loss on the SB already drastically alters the overall wind
evolution from the beginning and inhibits the evolution towards
asymptotic wind solutions. 

\section{Results}
\label{s.res}

\subsection{Mass and energy losses}
\label{s.loss}

The inner boundary values of the galactic wind simulations will be
influenced by the mass and
energy losses of the SB (see~Eqs.~\ref{e.msb} and \ref{e.ene_sb}) 
through the area $A_0$ as sketched in 
Fig.~\ref{f.ini_mod}. Taking a typical value of $10^{-3}\Msol\,{\rm yr^{-1}}$ we 
can estimate that over a time interval of $10^8\,$years about $10^{5}\Msol$
are transported out of the SB. This mass is comparable to the 
initial total mass of the SB and this mass loss decreases the gas
density $\rho_0(t)$. The outflow of material becomes easier because less material
has to be accelerated out of the galactic potential.  

In addition to this density change, the energy losses 
reduce the thermal pressure as well as the CR pressure to drive the flow.
We emphasize that CRs are scattered by the magnetic
fluctuations running outwards at the Alfv\'enic speed $\vA$ 
and therefore the CRs are advected outwards with the combined speed of $u_0+\vA$. 
The diffusion of particles with a mean diffusion coefficient $\bk$ \citep[see e.g.][]{Dru}
modifies the CR gradients and adds to the losses of energetic particles as formulated
in Eq.~(\ref{e.flux}). The range of this mean diffusion coefficient $\bk$ is expected to
be in the range between $10^{28}\,{\rm cm^2\,s^{-1}}$ and $10^{30}\,{\rm cm^2\,s^{-1}}$ 
\citep[e.g.][]{Axf:81}. The acceleration of energetic particles is discussed in more
detail in Sect.~\ref{s.acc}.

\begin{figure}   %                                                         Fig. 5
%
% inner boundary with back reaction, small scale
%
% \resizebox{\hsize}{!}{\includegraphics{Plots_Paper_v2/plot_IBC_ML_paper_K28.eps}}
  \resizebox{\hsize}{!}{\includegraphics{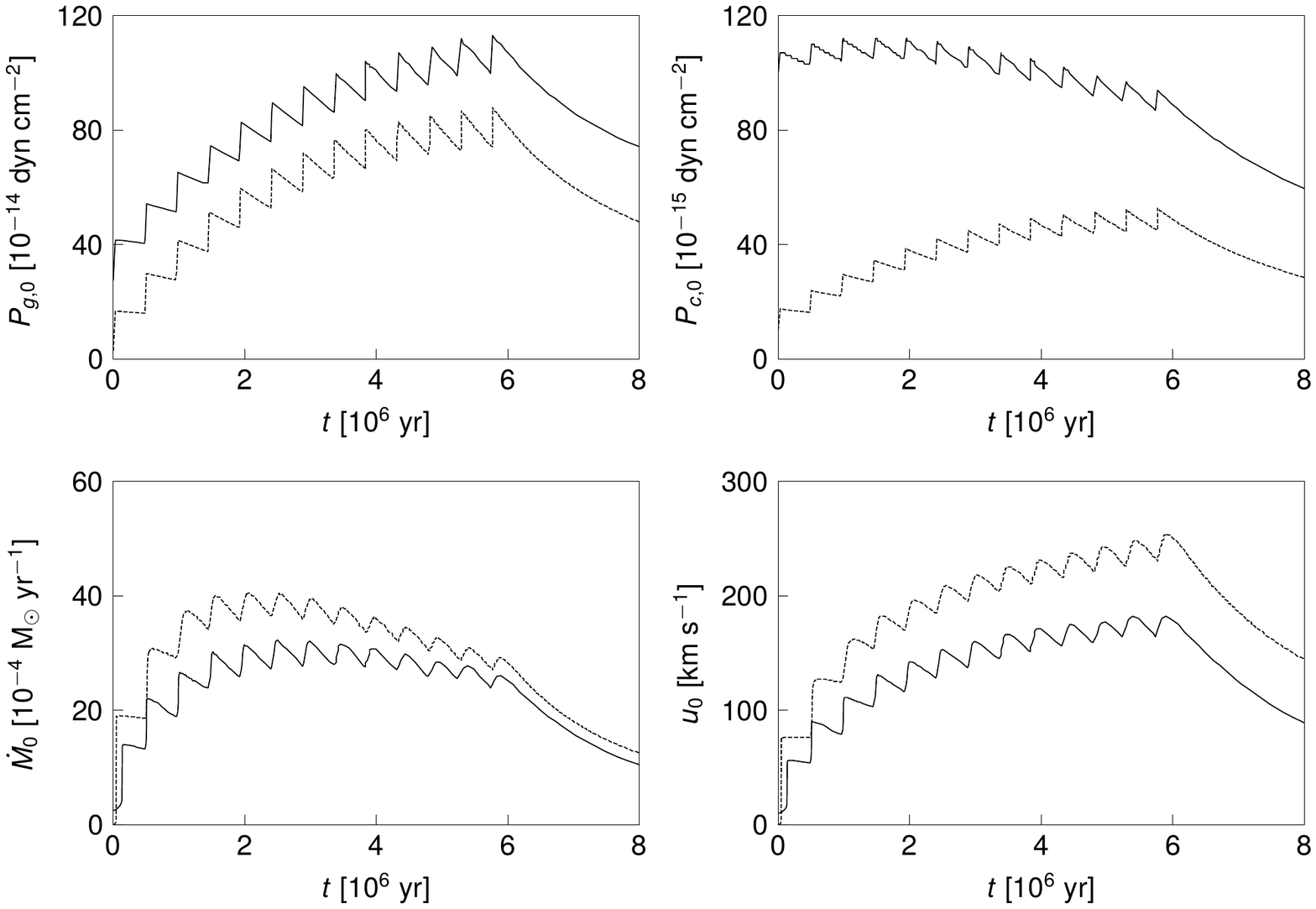}}
  \caption{Gas pressure $P_{{\rm g},0}$ (upper left panel) at the inner boundary ($z_0=1\,{\rm kpc}$)
           of the flux tube for {\sf Model W} (full line) and {\sf Model H} (dashed line) as
           a function of time including the feedback by galactic wind loss. The CR pressure
           $P_{{\rm c},0}$ (upper panel right) is lower by about a factor of $10$ compared to the gas pressure. 
           All boundary values decrease between the SN explosions. After $6\!\cdot\! 10^6{\rm yr}$, 
           when the last SN exploded, the boundary values get monotonically smaller, caused by the
           outflow from the SB. The lower panel (left) exhibits the mass loss 
           by the galactic wind, which reaches a maximum rate around 
           $3.2\cdot 10^{-3}\Msol\,{\rm yr^{-1}}$
           for {\sf Model W} and $4.1\cdot 10^{-3}\Msol\,{\rm yr^{-1}}$ for 
           {\sf Model H}. In both models the maximum outflow velocity $u_0$ (lower panel right) is reached at
           the last SN explosion.}      
  \label{f.ini_val_mod}
\end{figure}

Both effects become clearly visible in Figs.~\ref{f.ini_val_mod} and 
\ref{f.ini_val_mod_global} (full lines for {\sf Model W}, dotted lines for {\sf Model H})
where the time-dependent inner boundary values of the thermal pressure $P_{{\rm g},0}(t)$ and 
the CR pressure $P_{{\rm c},0}(t)$ are plotted as a function of time.
The variations of the gas density $\rho_0(t)$ and the velocity $u_0(t)$ 
determine the mass loss $\dot M_0(t)$ streaming through the flux tube. 
Comparing
Fig.~\ref{f.ini_val_mod} to the behaviour of the boundary values without
the feedback in Fig.~\ref{f.ini_val_mod_no} we clearly see the decreasing pressures
between the individual SN explosions at every $\dtsn=4.8\cdot 10^5\,$years. After
the last SN at $6\cdot 10^6\,$years all boundary values decrease monotonically. The
gas pressures increase until the last SN explodes but the smaller cosmic
ray pressures (upper panel, right) are affected by diffusive losses and reach
their maximum at $1.44\cdot 10^6\,$years ({\sf Model W}) and 
$5.76\cdot 10^7\,$years ({\sf Model H}), respectively. From the dynamical
point of view the CR pressure is less important at the inner boundary and
is typically a factor of $10$ below the thermal pressure (cf.~the different scales in
Fig.~\ref{f.ini_val_mod}).
Since in {\sf Model H} the large gradients lead to an overall lower background wind density,
the bubble more rapidly loses the mass owing to higher mass loss rates 
(Fig.~\ref{f.ini_val_mod}). We find a maximum mass loss rate of 
$4.1\cdot 10^{-3}\Msol\,{\rm yr^{-1}}$
({\sf Model H}) at time $1.96\cdot 10^6\,$years and $3.2\cdot 10^{-3}\Msol\,{\rm yr^{-1}}$
at time $2.46\cdot 10^6\,$years ({\sf Model W}).
No stationary solutions can develop and the variations at the inner boundary 
will terminate the mass and energy loss owing to 
 realistic feedback from the wind.

\subsection{Long term evolution}
\label{s.long}

Figure~\ref{f.ini_val_mod_global} exhibits the change of the inner boundary
values if we take into
account the mass and energy losses of the SB through a galactic wind. 
Over the timescale of $10^8\,$years, we see that after $10^7\,$years  
the mass loss $\dot M_0$ ceases (lower panel, left). As already seen in Fig.~\ref{f.ini_val_mod},
the inner boundary values of $P_{{\rm g},0}$, $P_{{\rm c},0}$, $\dot M_0$ , and
$u_0$ start to decrease immediately after the 
last SN explosion and reduce thereby the ability to drive an outflow over
longer timescales. The effect of mass and energy losses can be seen by comparing the
temporal evolution of the galactic wind without the feedback as plotted in 
Fig.~\ref{f.ini_val_mod_no_global}. 
After less than $25\cdot 10^6{\rm yr}$ the outflow has ceased for both models and
no more material can be transported outwards into the extragalactic medium.
For $t>25\cdot 10^6{\rm yr}$ we find almost no more variations of the
thermal pressure $P_{{\rm g},0}\simeq 2\cdot 10^{-13}\,{\rm dyn\,cm^{-2}}$. 
The CR pressure $P_{{\rm c},0}$ is steadily decreasing after the end
of the SN explosions mediated by diffusive losses. We see
from the lower panels of Fig.~\ref{f.ini_val_mod_global} that larger
pressures at the inner boundary do not automatically lead to higher mass
loss rates, by comparing {\sf Model W} and {\sf Model H}. The almost
hydrostatic background structure of {\sf Model H} enables a faster and more 
massive outflow.

We note that, as opposed to models without feedback, the
mass loss of {\sf Model H} remains higher than {\sf Model W} during the
whole wind evolution, because the
solutions have no possibility to develop into a stationary wind.  
%%
%% DB: vielleicht eine kurze Erklärung hier? Bitte Zeitskalen prüfen.
%%
This can be easily seen if one compares the time the flow takes to reach the outer boundary at $300\, {\rm kpc}$ of about $2.5 \, 10^7$ years with the stationary flow time $\int dz/u(z)$, which was found to be approximately $10^8$ years.

\begin{figure}   %                                                         Fig. 6
%
% inner boundary with back reaction, large scale
%
% \resizebox{\hsize}{!}{\includegraphics{Plots_Paper_v2/plot_IBC_ML_paper_K28_wide.eps}}
 \resizebox{\hsize}{!}{\includegraphics{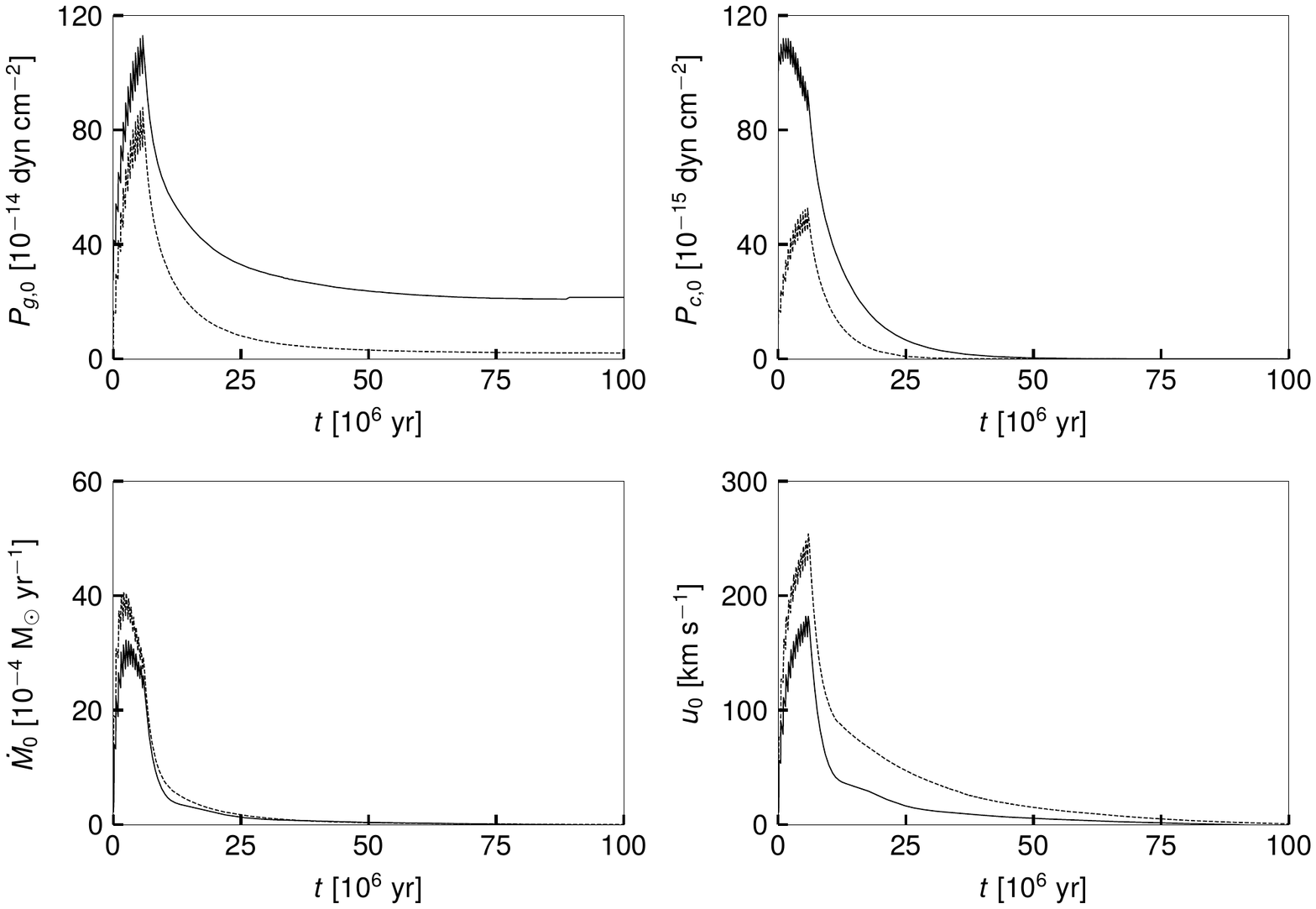}}
  \caption{Gas pressure $P_{{\rm g},0}$ (upper panel, left) at the inner boundary ($z_0=1\,{\rm kpc}$)
           for {\sf Model W} (full line) and {\sf Model H} (dashed line) as seen over
           a time interval of $10^8\,$years for the case when mass and energy losses
           are included. The CR pressure $P_{{\rm c},0}$ (upper panel, right)
           is about a factor of $10$ less than 
           the gas pressure (note the change of scale).
           The decrease of the mass loss rate $\dot M_0$ (lower panel, left) occurs
           already $10 \cdot 10^6{\rm yr}$ after the SNe have triggered an
           outflow. After this time the galactic wind changes its nature as inferred
           from the decreasing values of the outflow velocity $u_0$ (lower panel, right).}   
  \label{f.ini_val_mod_global}
\end{figure}

The global temporal evolution of such galactic winds can be discussed in the
context of Fig.~\ref{f.int_loss}, where integrated quantities are plotted
over $10^8\,$years and in less than $25\cdot 10^7\,$years all integrated
quantities reach their final values. According
to our initial conditions we put $0.9 \cdot 25\Esn$ as thermal energy
and $0.1 \cdot 25\Esn$ as CR energy into our SB located beneath the 
galactic wind. The upper panels of Fig.~\ref{f.int_loss} plot the integrated loss
of thermal energy $E_{{\rm th},0} = \int \dot E_{\rm th}\,dt$ as well as the integrated 
advected CR energy in units of the SN energy of $10^{51}\,{\rm erg}$. The
removed thermal energies are almost identical for {\sf Model W} (full lines) 
and {\sf Model H} (dashed lines). Since CRs can leave the SB through
advection $E_{{\rm CR, adv},0}$ by a velocity $u_0+\vA$ , as well as through a diffusive flux 
$E_{{\rm CR, diff},0}$, the left panels of Fig.~\ref{f.int_loss} plot both contributions
separately as described in Eq.~\ref{e.flux}. Comparing these figures we conclude that 
the diffusive contribution $E_{{\rm CR, diff},0}$ at the inner boundary 
remains small compared to the streaming of CRs $E_{{\rm CR, adv},0}$, 

From the integrated mass losses $\int \dot M(t)\,dt$ as presented in Fig.~\ref{f.int_loss},
we get $3.3\cdot 10^4\Msol$ for {\sf Model W} and  $2.6\cdot 10^4\Msol$ for {\sf Model H} 
characterized by the lower initial external density. We find that both models 
transport almost $25\,\Esn$ of thermal energy into the intergalactic
medium. {\sf Model W} (full lines) releases almost the same amount of energetic particles, 
whereas in {\sf Model H} (dotted lines) the CRs contribute only an 
equivalent of about $10\, \Esn$ to the extragalactic space.
% 
%%
%% DB: Fig. Caption etwas geändert
%%
\begin{figure}   %                                                         Fig. 7
%
%               integrated quantities
%
% \resizebox{\hsize}{!}{\includegraphics{Plots_Paper_v2/plot_IBC_integrated_ML_paper_K28_wide.eps}} 
 \resizebox{\hsize}{!}{\includegraphics{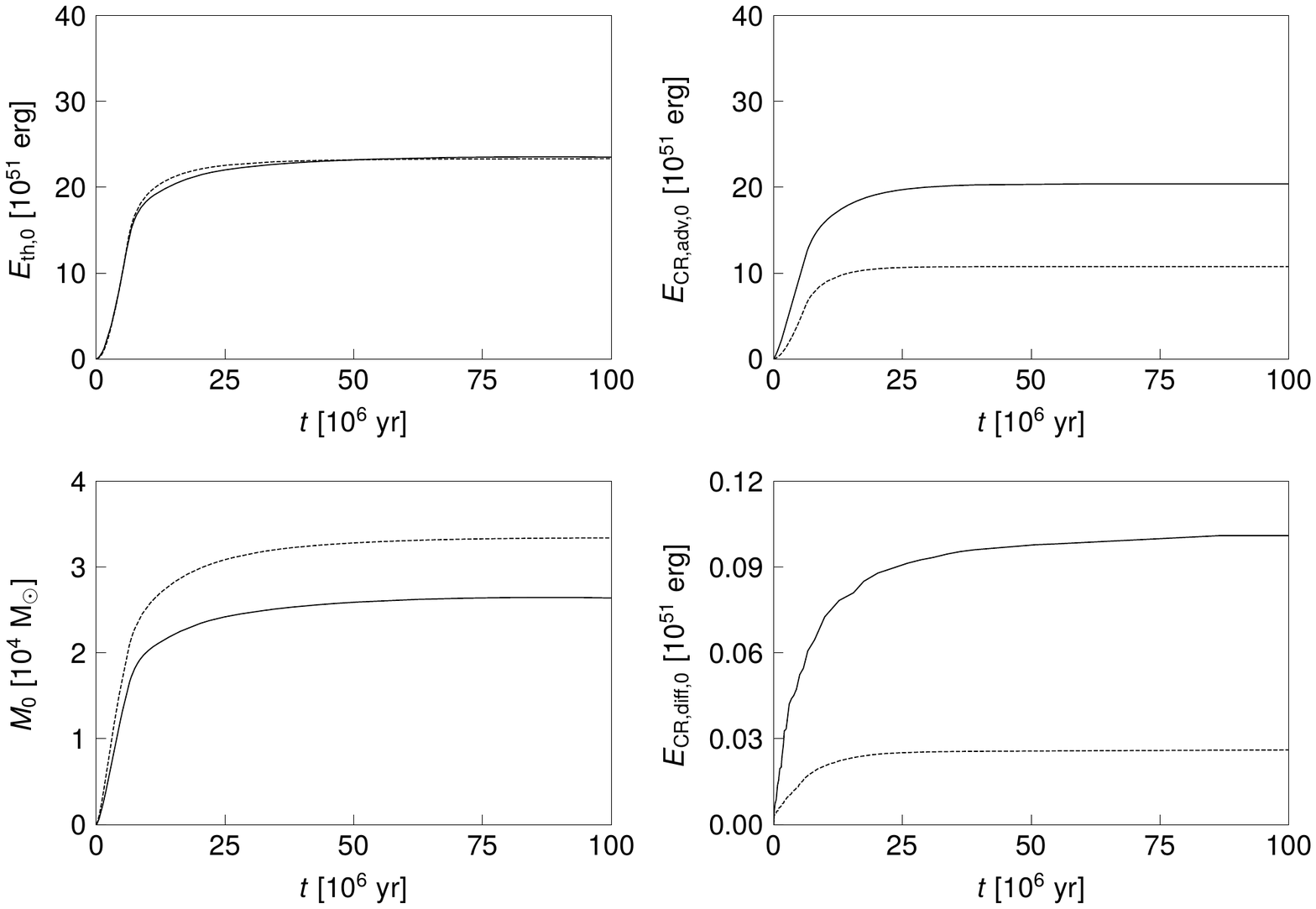}}
  \caption{Energy loss $E_{\rm th,0}(t)$ in units of $\Esn$ (upper left panel), 
           the CR energy loss contributions due to advection (upper right), the integrated mass loss $M_0(t)$ 
           in units of $10^4$ solar masses $\Msol$ (bottom left), and
           $E_{\rm CR,adv,0}$ and diffusion $E_{\rm CR,diff,0}$ in units of $\Esn$ (bottom right) as a function of time.
  }
  \label{f.int_loss}
\end{figure}

%%
%% DB: hier fehlt noch eine Zahl; wie wäre es mit P_{cr} ~ K (gamma_c -1) E_{th}, wobei K = 0.01 - 0.1 ist;
%% Im SNR gilt E_{cr} ~ 0.1 E_{th}; solange SNe explodieren wird K ~ 0.1 eine gute Näherung sein, wenn aber die Energiequelle 
%% erlischt, wird K ~ 0.01 wahrscheinlicher sein, da  in der SB fast kein Magnetfeld ist, und die Teilchen schnell entweichen 
%% und erst im Halo wieder stärker gestreut werden.
%%
% EAD:  Ein unterer grenzwert wird nicht erreicht, aber sonst gilt   
%  The CR pressure has been set to a lower limit of $P_{\rm c, SB} = 10^{-18}\,{\rmdyne\,cm^{-2}}$ inside the SB. 

Taking a typical lateral dimension of $b=251\,{\rm pc}$ at the outbreak 
(see Sect. \ref{s.komp})
and adopting a lower level of $\bk=10^{28}\,{\rm cm^2s^{-1}}$ , we obtain a typical timescale
of $\tau = b^2/\bk \simeq 2\cdot 10^6\,{\rm years}$ for the diffusion of energetic particles into
the SB. This timescale is short compared to the typical evolution
timescale of the galactic wind. We conclude that the CR pressure inside a SB will not 
decrease under the mean galactic cosmic pressure, assumed to be $10^{-18}\,{\rm dyn\,cm^{-2}}$ for our wind models. 
%%
%% DB: das ist eine interessante Fragenstellung; wenn ich es richtig verstehe, möchtest Du zeigen, dass die CR_Energiedichte nicht unter den ISM-Mittelwert fallen kann
%% Dieser Mittelwert hängt allerdings vom mittleren Magnet- und Wellenfeld ab, welche kappa bestimmen; in der SB sind beide niedrig, so dass theoretisch infolge von Gas- Wellen und CR-Evakuierung der Wert tatsächlich niedriger sein könnte; aus meiner Sicht ist die Frage daher nicht so einfach zu beantworten
%%
This fact explains why in Fig.~\ref{f.int_loss} the integrated values of the 
CR energy losses by advection $E_{\rm CR,adv,0}$ and by diffusion $E_{\rm CR,diff,0}$
can reach values comparable to the thermal losses $E_{\rm th,0}(t)$. 
%%
%% DB; wobei der Diffusionsverlust nur 5% des Advektionsverlustes beträgt
%%
Nevertheless we point out that
the main drivers of the flow are the thermal pressure gradients and that CRs 
play only a minor role in initiating the outflow. From the lower right panel of
%% DB: ich habe oben, hier und weiter unten alle Typos ohne weitere Kommentare verbessert
Fig.~\ref{f.int_loss}, where the integrated losses $E_{\rm CR,diff,0}$ caused by the diffusion of energetic
particles are plotted as a function of time, we infer that the diffusive losses out of the bubble are
negligible compared to the flux of CRs $E_{\rm CR,adv,0}$ transported out by a galactic wind,
and CR gradients are therefore not responsible for driving the outflow.   
%%
%% DB: eigentlich ist kappa = 10^28 eher niedrig, da die Diffusionslänge L = kappa/v_A ~ 10^28/10^7 ~ 10^21 cm ist, wobei v_A die Alfvengeschw. und damit die mittlere 
%%  Diffusionsgeschw. der Teilchen infolge Streuung ist.
%%
Due to the large mean CR diffusion coefficients of $\bk\ge 10^{28}\,{\rm cm^2s^{-1}}$ , only gradients
on the order of several kilo parsecs will be affected through diffusive effects. 
Consequently, at distances beyond $30\,{\rm kpc}$
CRs become increasingly important to further accelerate the material 
(see e.g.~Fig.~\ref{f.rad_loss_W}), provided that reasonably strong coupling to the plasma is ensured.
%%
%% DB: hängt von der Kopplung ab
%%

\subsection{Coalescence of shock waves}
\label{s.coal}

The variations of the inner boundary conditions will mainly affect the region above the
SB, which in our models is the region above $1\,{\rm kpc}$. Every SN explosion generates
non-linear waves, which travel into the previously shaped medium. To illustrate these
interactions in Fig. 8 we have plotted 
the location of some shock waves for {\sf Model W} and {\sf Model H} triggered by the pressure jump
from a SN explosion within the SB. The time axis is given in units
of mean explosion interval of $\dtsn = 4.8\cdot 10^5{\rm yr}$. These plots are generated
for the original computational data, and an adaptive grid \citep[]{DD} with up to $2000$ 
grid points has been used 
to follow the non-linear evolution of these waves. After each SN explosion a new shock is
initiated at the inner boundary located at $1\,{\rm kpc}$ and the moving grid points
concentrate there to resolve the new emerging features. As summarized in Table~\ref{t.shock},
the developing shocks start with a rather small Mach-number and hence the adaptive grid has
to react to very tiny changes of physical variables. 

From the physical point of view the interactions of these forward and reverse shocks cause
rather complicated flow structures even for the case of pure gas dynamics \citep{CF}. The
overtaking of shock waves results in a transmitted shock,  reflected in general in a weak rarefaction
wave and a contact discontinuity between them. Two head-on colliding shock waves are weakened, 
 retarded, and penetrate each other. Between them they leave a contact discontinuity
within an expanding zone of constant pressure and flow velocity. All these structures will
emerge on the non-constant background of our initial models. As we use an adaptive grid,    
the grid points rush towards such interaction zones if new features are generated.
In addition the variations at the inner boundary have to be resolved and these
grid motions can generate small disturbances in the determination of the shock location
seen as kinks in outer shock locations. Owing to all these new structures, 
the exact location of a
particular shock wave on a discrete grid is difficult to determine and
the paths plotted in Fig.~\ref{f.shocks} are computed afterwards 
and have no influence on the shock propagation itself. We emphasize that
we have adopted a conservative numerical scheme. Moreover the 
later shocks travel almost at sound velocity,~$M\simeq 1$ as
seen in Table~\ref{t.shock}, and their small pressure jumps are
therefore difficult to localize. 

\begin{table}
\caption{Pressure changes and Mach numbers of the first five shock waves.}
\label{t.shock}
\centering
\begin{tabular}{l|r|c}
 \toprule
 Shock number  & {\sf Model W}   & {\sf Model H} \\
 \midrule
  1 &          $ p_1/p_2 = 1.50 $ &  $= 6.05$ \\
    &          $ M = 1.18$        &  $= 2.24$ \\                
  2 &          $ p_1/p_2 = 1.33 $ &  $= 1.83$ \\
    &          $ M = 1.13$        &  $= 1.29$ \\ 
  3 &          $ p_1/p_2 = 1.25 $ &  $= 1.46$ \\
    &          $ M = 1.09$        &  $= 1.17$ \\ 
  4 &          $ p_1/p_2 = 1.20 $ &  $= 1.31$ \\
    &          $ M = 1.08$        &  $= 1.12$ \\
  5 &          $ p_1/p_2 = 1.27 $ &  $= 1.23$ \\
    &          $ M = 1.06$        &  $= 1.09$ \\                                                  
 \bottomrule
\end{tabular}
\end{table}

Table~\ref{t.shock} shows that the shock strength decreases as more and more
SN explode because the SN energy added to the SB decreases relative to
the total energy already deposited within the bubble. The typical Mach numbers
are around $M\simeq 1$ for {\sf Model W,} where wind from the SB 
evolves into a background of an earlier wind solution. 
%%
%% DB: Ergänzung
%%
For {\sf Model H}
starting from an almost hydrostatic background the Mach numbers are slightly higher, especially for the first shock that moves into a static medium, whereas later shocks will propagate into a medium set into motion by previous shock waves, thereby converging towards sound waves with $M\simeq 1$.  

%
%  shock waves at the innermost regions
% 
\begin{figure}   %                                                         Fig. 8
%
%%
%% DB: eigentlich ist space time x-t, während wie hier t-z time-space wäre oder man vertauscht tatsächlich die Achsen wie in der SR
%%
% \resizebox{\hsize}{!}{\includegraphics{Plots_Paper_v2/plot_timeshocks_ML_paper_K28_detail.eps}}
  \resizebox{\hsize}{!}{\includegraphics{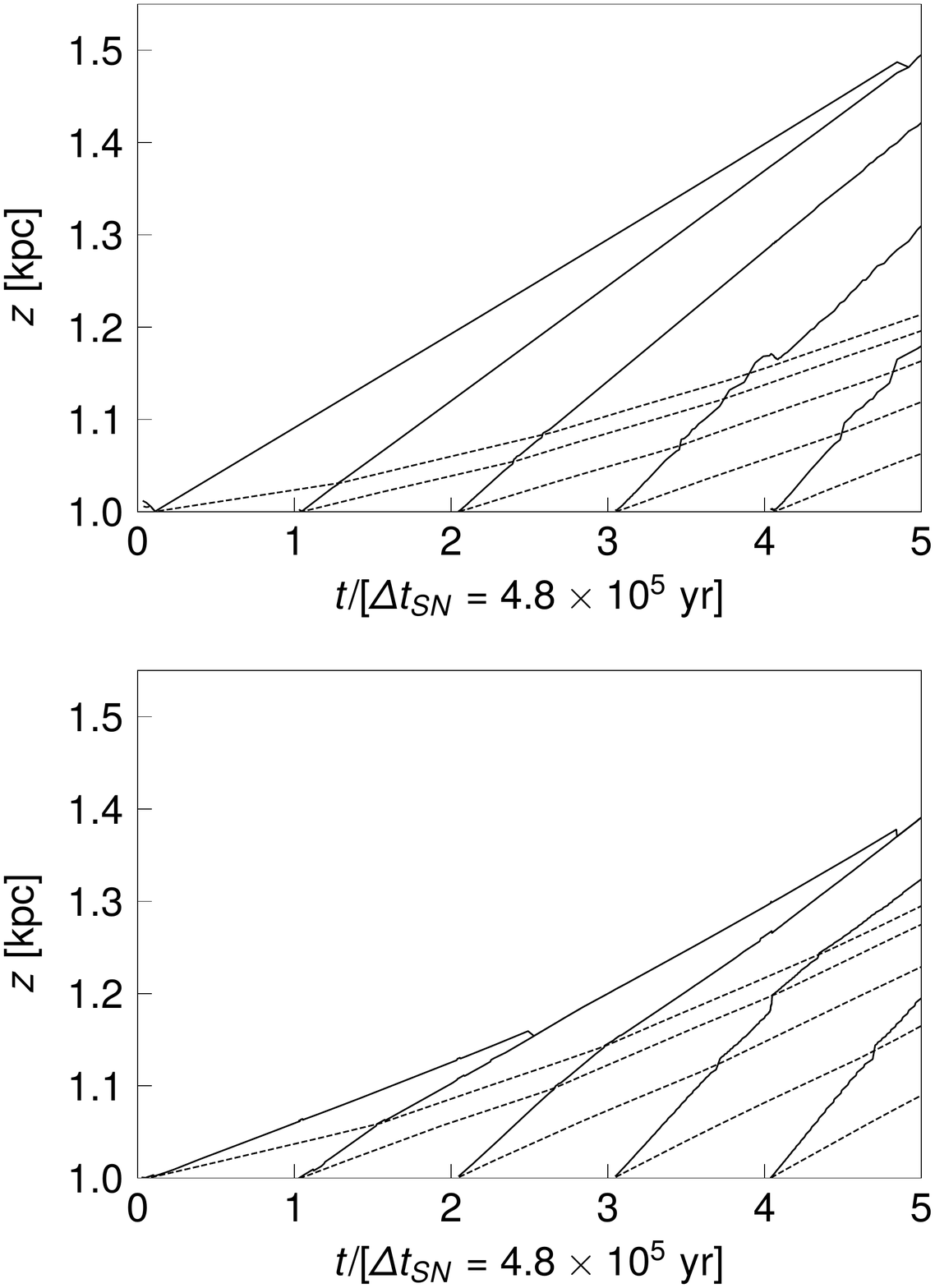}}
  \caption{Coalescence of the first five forward shock waves (full lines)
           in a time-space diagram for {\sf Model W} (upper panel) and {\sf Model H} (lower panel) at
           the innermost $1.5\,{\rm kpc}$. Dotted lines remark the reverse shocks.
           The time is given in units of the SN explosion interval $\dtsn$. The reverse
           shocks (dotted lines) move at smaller speeds and collide with the forward
           shock, slowing them down.}
  \label{f.shocks}
\end{figure}

%% DB: "inclination" statt "upward bending"
A close inspection of Fig.~\ref{f.shocks} verifies the aforementioned features of shock interactions. 
From the inclination of the path of the outermost forward shock we see that the later shock 
waves are running faster and overtake the earlier shock waves. The head-on collision of a 
forward shock (full line) with a reverse shock (dotted line) 
always causes a slowdown of the forward shock, visible as flatter paths. 
The corresponding reverse shock gets accelerated as seen by steeper paths. 
%%DB: ein Satz
The effects of acceleration and deceleration are most pronounced when the Mach number of the respective shocks is high, whereas for weaker (and hence later) shocks the effect is barely visible in Fig.~\ref{f.shocks}.
These features are located in the plane part of the flow where we can neglect geometrical effects 
because the spatial scale of Fig.~\ref{f.shocks} is limited to $1.5\,{\rm kpc}$, 
small compared  to the opening scale of the flux tube $z_b = 15\,{\rm kpc}$.  

The action of three pressure gradients resulting from 
thermal gas, CRs, and Alfv\'enic waves 
enhances the complexity of these numerous non-linear wave interactions. 
The shock waves accelerate energetic particles through a first-order Fermi process and the resulting 
CR pressure gradients modify the shock structure \citep[e.g.][]{Dru:Voe} by generating
a CR precursor through diffusion. 
This whole zone of interacting non-linear waves propagates outwards 
along the pressure gradients and gets expanded. Only two shock waves survive and 
in Table~\ref{t.coal} we summarize the 
evolution of this transitional zone by the typical time and distance after 
which all waves have been merged. Afterwards the flow structure is  dominated by a 
single outwards-moving strong shock wave followed by a reverse shock that is also advected outwards.

\begin{table}
\caption{Typical distances, times, and Mach number where the
         different shock waves merge into a single propagating forward shock.}
\label{t.coal}
\centering
\begin{tabular}{l | c| c}   %$1.67\!\cdot\!10^{-27}$
\toprule
                      &  {\sf Model W}  & {\sf Model H} \\
\midrule
 $t_{\rm merge}\,[{\rm yr}]$      &  $1.5 \cdot 10^{7}$    & $1.0 \cdot 10^{7}$  \\
 $z_{\rm merge}\, [{\rm kpc}]$    &  $7$                   &   $5$ \\
 Mach number                      &  $2.4$                 & $10.6$ \\
 \bottomrule
\end{tabular}
\end{table}

We emphasize that these shock waves coalesce into a single wave 
rather close to the galactic plane at radii given in Table~\ref{t.coal}. 
As inferred from Fig.~\ref{f.shocks}, the first two shock waves already merge 
at $1.6\,{\rm kpc}$ in the case of {\sf Model W} and at $1.2\,{\rm kpc}$ for {\sf Model H}.
For {\sf Model W} the overall merging distance is at $z_{\rm merge,W} = 7\, [{\rm kpc}]$ 
corresponding to $1.5 \cdot 10^{7}\,$years and after this event the overall evolution
is basically characterized by a single forward shock wave travelling into the 
intergalactic medium. The case of an almost hydrostatic initial configuration ({\sf Model H}) 
leads to a faster shock merging 
at $z_{\rm merge,H} = 5\, [{\rm kpc}]$ after $10^{7}\,$years because the steeper gradients 
accelerate the coalescence of the different shock waves (see also Fig.~\ref{f.shocks}). After the flow features have reached a distance larger than 
$z_{\rm merge}\, [{\rm kpc}]$ a simpler, single shock approximation 
is suitable to describe the galactic wind and the results can directly be compared 
with the previous time-dependent wind models of \citet{DB1}.

The position-time diagram of Fig.~\ref{f.shocks} illustrates
the evolution of the merged forward and 
reverse shock over the time interval of $10^8\,$years. The thin lines correspond to
an evolution of the inner bubble without the feedback from galactic wind losses and
are plotted mainly as reference to the simpler models in \citet{DB1}. 
Clearly, the more realistic models with feedback (thick lines) exhibit a similar evolution
at small timescales and the difference already becomes pronounced after $10^7\,$years. 
Since the loss term decreases the pressure values at the base of the flux tube, the overall
evolution of all waves is characterized by smaller velocities. Again, we can
distinguish a merging period of $t\le \,1.5\cdot10^{7}$years for {\sf Model W} 
and $t\le \,1.0\cdot10^{7}$years for {\sf Model H} followed by an expansion on larger scales
at almost constant shock speeds. The full lines indicate the location of the forward shock, the 
dashed lines point at the reverse shocks.

In {\sf Model W} with feedback (lower lines) we find a finite expansion time 
for the reverse shock, which reaches a maximum spatial dimension of $z_{max} = 5.43\,{\rm kpc}$
(see also Fig.~\ref{f.rad_loss_W}).  
The region interior to the reverse shock has been depleted and depressurized in such a way 
that the reverse shock starts to propagate inwards and reaches our inner boundary at
$1\,{\rm kpc}$ after $1.94 \cdot 10^8\,$years, where our computation ends. At this time
the outer shell is located at a distance of $92\,$kpc.  The
forward shock propagates at a speed of $\us=490\,$km/s. 

For {\sf Model H} the structures can propagate to large distances because 
the initial model exhibits steeper gradients and less material has to be removed from the flux tube.
Moreover the steeper density gradients lead to a higher shock acceleration, as during shock propagation progressively less material has to be compressed by the shock according to the equation of continuity.
%%
%% DB: wenn die Masse in der Flussröhre für W- und H-Modelle gleich wäre, sollten die W-Modelle eigentlich Schocks haben, die weiter nach außen propagieren, da der Wind 
%% ja in dieselbe Richtung läuft wie die Schocks oder? Andererseits sind die Dichtegradienten in den H-Modellen steiler, was die Schocks beschleunigt, was zu höheren 
%% Geschwindigkeiten führt. Ich habe daher einen Satz eingefügt. 
%%
Again, the models including the wind losses (thick lines) 
exhibit lower shock speeds, that is,~the asymptotic speed of the forward shock decreases from
$\us = 2980\,$km/s to $2770\,$km/s. The corresponding values for the reverse shock are subject
to larger variations; the speed drops from
$980\,$km/s to $390\,$km/s. After about $10^8\,$years the forward shock has reached our outer
boundary located at $300\,$kpc.
%%
%% DB: müsste es beim reverse shock nicht umgekehrt sein, d.h. er wird in den H-Modellen langsamer, da er in einen positiven Dichtegradient hineinläuft?
%% In der Fig. 8 müsste wohl Model A = H und Model B = W sein; dann würde man genau das sehen. Ich habe daher einen Satz dazu geschrieben.
In contrast, and as can be seen in Fig.~\ref{f.shocks}, the reverse shock has a lower speed in {\sf Model H} because it is propagating into an ever-increasing density profile.

\subsection{Evolution of spatial structures}
\label{s.spatial}

The more detailed discussion of the spatial evolution 
is focused on models including mass and energy losses
by the galactic wind. Models without these back reactions are 
rather similar to those of \citet{DB1}. 

%%
%% DB: Satz war unvollständig; Ergänzung richtig?
%%
In Figs. \ref{f.rad_loss_W} and \ref{f.rad_loss_H} we plot the spatial evolution of the
galactic winds over more than $10^8\,$years for {\sf Model W} and {\sf Model H}. 
The most prominent feature is the
forward shock wave running at an asymptotic speed of $\us = 490\,{\rm km/s}$ down
the density gradient within the previous wind structure in {\sf Model W}, plotted at
times $7\cdot 10^7\,$, $1.4\cdot 10^8\,$ , and $2.1\cdot 10^8\,$years. 
%%
%% DB: den Wert von 490 km/s kann ich in Fig. 10 rechts unten nicht finden; meinst Du 390 km/s oder kann man das nicht ablesen, da zur Schockgeschw. noch die upstream-
%% Geschw. dazu gerechnet werden muss?
%%
The reverse shock is less important but becomes more visible as 
it starts to move inwards after about $8\cdot10^7\,$years. 
%%
%% DB: sage mir ob die nachfolgende Beobachtung richtig ist, falls nicht, nehme den Satz wieder heraus; ich sehe in Fig. 10 unten, dass sich der reverse shock von 1.4 - 2.1 
%%      10^8 yr nicht mehr bewegt
%%
It is interesting to observe that the reverse shock transforms into a so-called termination shock, known from stellar winds, 
at around $2 \cdot 10^8\,$years, when it becomes almost stationary in the Eulerian frame of reference. 
Once the SB flow at the distance of this shock vanishes, it will move towards the galactic disc, thereby 
reheating the flow in the inner flux tube, similar to a supernova explosion. 

Comparing the three pressure contributions (upper panels of Figs.~\ref{f.rad_loss_W} and 
\ref{f.rad_loss_H}), the CR pressure gradients will dominate the flow 
above $20\,$kpc. Throughout the whole evolution the increase of CRs in the
post-shock region, as well as the precursors in front of
the forward shock, exhibit the acceleration of energetic particles in these
large scale wind structures. However, due to Mach numbers of $M\simeq 3,$
the overall acceleration efficiency remains 
so low that the CR pressure is influenced 
by diffusive losses towards the inner flux tube, which leads to steep gradients 
(cf.~upper right panel). The
values at the inner boundary (cf.~Fig.~\ref{f.ini_val_mod_global}) show that after $5\cdot 10^7\,$years
the CR pressure remains extremely small and the outflow
velocity $u_0$ decreases to a few kilometers per second. The spatial scale is roughly given by $l=\bk/u,$ and with
$\bk=10^{28}\,{\rm cm^2\,s^{-1}}$ and $u\simeq 3\,$km/s we obtain $l\simeq 10\,$kpc
as also deductible from the plots. 
The scale reduced by a factor of $10$ for the wave pressure $\Pw$ (dashed lines, right upper panel)
illustrates that the waves are only a minor contribution to drive the flow outwards.
%%
%% DB: warum? in den stationären Windmodellen, liegt P_w sogar immer leicht über P_c
%%

The resulting galactic wind significantly reduces the energy 
content of the underlying SB
(see also Figs.~\ref{f.ini_val_mod} and \ref{f.ini_val_mod_global}, thick lines) already 
after the last SN has exploded, at~$t>6\cdot 10^6\,$years. At
some point the inner part of the flow is no longer supported by the 
various pressure gradients and starts to fall back for radii smaller than $30\,{\rm kpc}$.
As mentioned, the CR gradient is the main cause of the reversal of the flow direction.
The density structures reveal the propagating forward shock as well as the development of
a reverse shock inside $30\,$kpc when the wind
velocity vanishes and the material starts to fall inwards. Clearly, these flow features
are more clearly visible in the corresponding velocity plot (lower right panel) 
and after $10^8\,$ years we find material falling inwards with negative flow
speeds of more than $200\,$km/s. 
We stopped these computations when the reverse shock reached the inner boundary at
$2.12\cdot 10^8\,$years.  
%%
%% DB: ich habe hier ein Verständnisproblem (s.o.), denn der reverse 
% shock bleibt praktisch bei 10 kpc stehen; oder rast er von 2.1 bis 2.12 10^8 J. von 10 bis 1 kpc?
%%
%
\begin{figure}   %                                                         Fig. 10
%
%   radial structures for model Wind
%
% \resizebox{\hsize}{!}{\includegraphics{Plots_Paper_v2/plot_radial_ML_paper_K28_ModelA_wide.eps}} 
  \resizebox{\hsize}{!}{\includegraphics{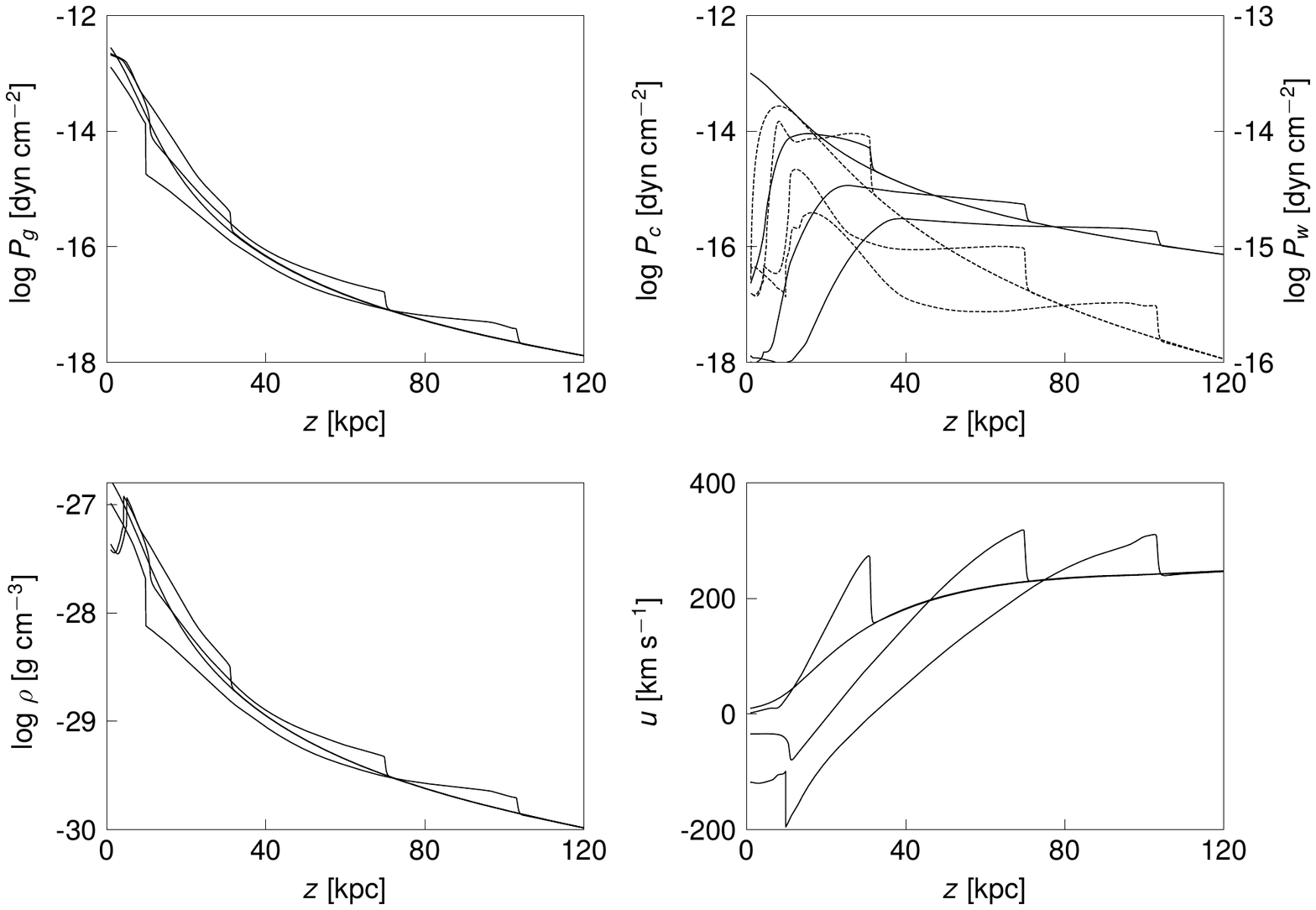}} 
  \caption{Radial structures for {\sf Model W} up to $120\,{\rm kpc}$ at three times,
   $7\cdot 10^7\,$, $1.4\cdot 10^8,\,$ and $2.1\cdot 10^8\,$years, exhibit the shock waves
   propagating along the flux tube. The upper panels show the different pressure
   contributions from thermal gas $\Pg$, CRs $\Pc,$ and waves $\Pw$ (dashed lines). 
   The density and velocity are characterized by the large scale outflow with decreasing 
   inner values. The last model at $2.1\cdot 10^8\,$years presents the reverse shock
   located around $11\,{\rm kpc}$ and the inner parts of the flow are directed inwards.}
  \label{f.rad_loss_W}
\end{figure}

In contrast to the wind background {\sf Model W,} 
the shock waves for {\sf Model H} with a low halo-like
density background travel 
as fast as $\us = 2780\,{\rm km/s}$ for the forward shock,  
and for the reverse shock, $390\,{\rm km/s}$ (see also Fig.~\ref{f.shocks}, thick lines).
Figure~\ref{f.rad_loss_H} exhibits the spatial evolution of the galactic wind for 
{\sf Model H} on the scale of $300\,{\rm kpc}$
plotted at three times, $t=3\cdot 10^7\,$, $6\cdot 10^7,\,$ and $9\cdot 10^7\,$years,
respectively. 
Again as seen in all panels, the most prominent feature 
is the forward shock travelling at an asymptotic
speed of $\us = 2780\,{\rm km/s}$. The reverse shock is located outside of the
denser material transported by the galactic wind with velocities of
a few $100\,$km/s (cf.~lower right panel). In {\sf Model H} this reverse shock gets
advected outwards with a speed of $390\,{\rm km/s}$. In the frame of the
contact discontinuity, the reverse shock is moving inwards as also inferred from the
growing distance between the two 'saw-tooth' features in the velocity (lower right panel).
Because of the
finite energy deposition by the SNe within the SB and the decreasing pressure support at
the inner boundary (cf.~Figs.~\ref{f.ini_val_mod} and \ref{f.ini_val_mod_global}, dashed lines), 
both shocks get weakened in time as they propagate down the density gradients. As
discussed in \citet{DB1}, this expansion phase yields to almost constant shock speeds. 
We terminate this calculation when the forward shock has reached the outer boundary ($300\,$kpc)
after $1.3\cdot 10^8\,$years. At this time the reverse shock is located at $71\,$kpc. Since we stop
the computations at this time, we cannot follow the further evolution of the reverse shock,
which at later times will also start to travel inwards.

\begin{figure}   %                                                         Fig. 11
%
%   radial quantities of model Halo
%
% \resizebox{\hsize}{!}{\includegraphics{Plots_Paper_v2/plot_radial_ML_paper_K28_ModelB_wide.eps}} 
  \resizebox{\hsize}{!}{\includegraphics{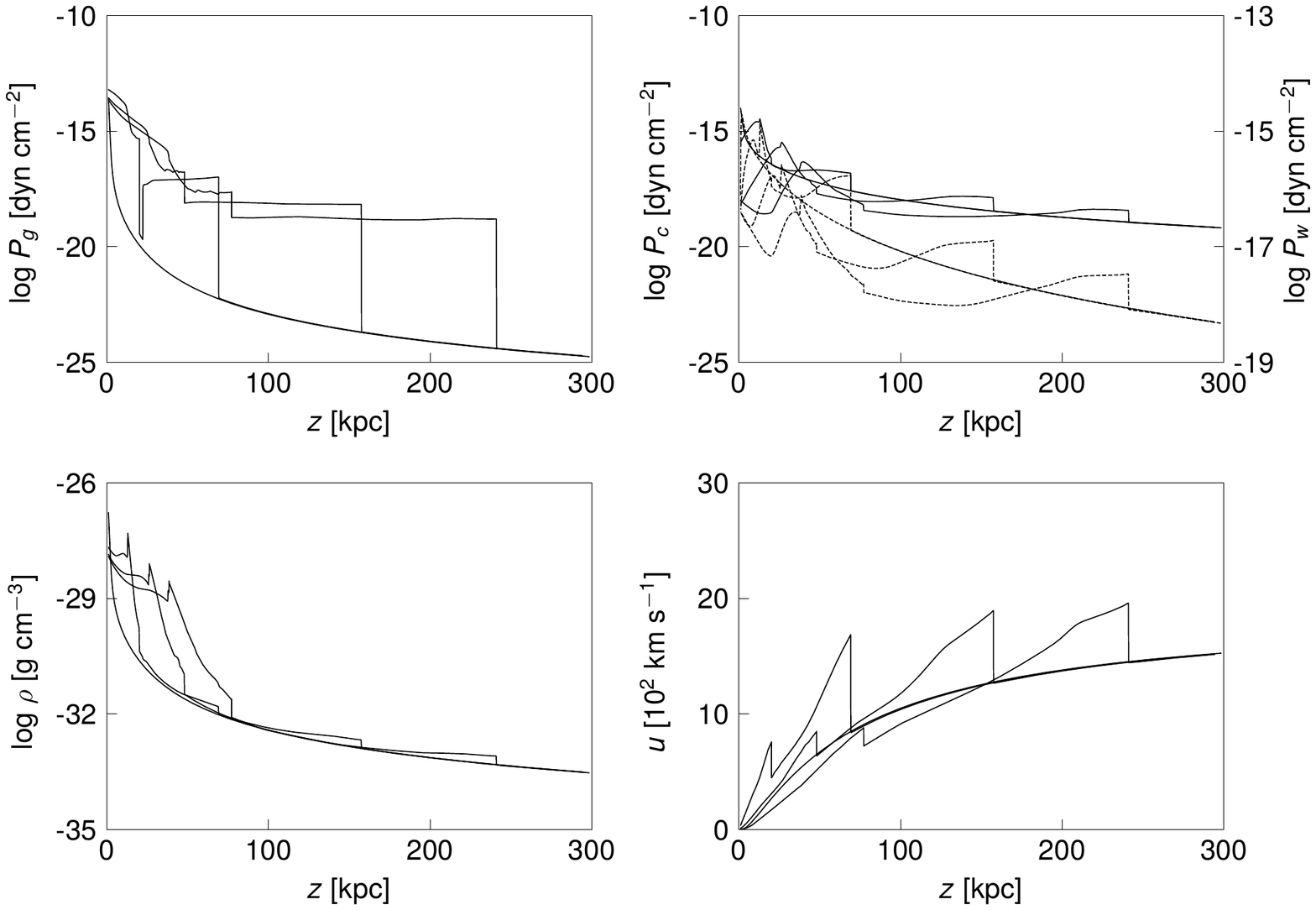}} 
  \caption{Radial structures for {\sf Model H}  on the large scale of $300\,{\rm kpc}$ at
   times of $3\cdot 10^7\,$yr, $6\cdot 10^7\,$yr, and $9\cdot 10^7\,$yr.
   The upper panels show the different pressure
   contributions from thermal gas $\Pg$, CRs $\Pc,$ and waves $\Pw$ (dashed lines).
   Flow velocites up to $2000\,{\rm km/s}$ are reached and the overall contribution of the
   reverse shock remains small.}
  \label{f.rad_loss_H}
\end{figure}

Although the gas pressure
dominates the flow around the inner boundary, the CR pressure provides
the major acceleration at larger distances around $30\,$kpc (upper panels
of Fig.~\ref{f.rad_loss_H}). The wave pressure (dashed lines) is typically 
below the particle pressure and therefore less important 
for these galactic winds. 
As inferred from Fig.~\ref{f.rad_loss_H}, the 
particles are mainly accelerated at the forward shock, the reverse shock plays only a 
minor role in the increase of the CR pressure. We emphasize that 
the presence of the diffusive shock acceleration (DSA) within galactic winds is 
described by a diffusive term in the energy equation of CRs, which 
together with $\gc=4/3$ leads to
an overall much smoother pressure structure over 
our computational domain of $300\,$kpc, which is~about two orders of magnitude in $\Pc$
compared to more than $10$ orders in thermal pressure $\Pg$. In contrast to {\sf Model W}
(presented in Fig.~\ref{f.rad_loss_W}) the lower background densities lead to a merged 
shock wave with a higher Mach number of $M\simeq 10$ (cf.~Table~\ref{t.coal}), which enables a
more effective particle acceleration. Therefore also the relative 
diffusive losses into the inner flux tube are smaller because the diffusive
length scale of $l=\bk/\us$ with $\bk=10^{28}\,{\rm cm^2\,s^{-1}}$ and 
$\us\simeq 3000\,$km/s is reduced to $l\simeq 10\,$pc, that is,~particle diffusion
remains a localized phenomenon. 

\subsection{Acceleration of cosmic rays}
\label{s.acc}

As inferred from Figs.~\ref{f.rad_loss_W} and \ref{f.rad_loss_H}, the pressure of the
bulk of energetic particles dominates the flow structure above $20\,$kpc in both cases.
Since we use a hydrodynamical description of the CRs, so-called CR hydrodynamics 
\citep[see e.g.][]{Dru} with $\gamma_c=4/3$, we have only limited information on the
maximum particle energy $E_{\rm max}$,  reachable during multiple encounters with
shock waves triggered by individual SN explosions. This
important pressure contribution emerges from  particles with lower energies. 
To obtain an estimate of the 
individual particle momentum, we can integrate
\begin{equation}    \label{e.pmax}
   \frac{d\pmax}{dt} = \frac{\pmax}{\tacc} \: ,
\end{equation}
where the diffusive acceleration timescale $\tacc$ is defined through
a shock wave with up-stream velocity $u_1$ and down-stream
velocity $u_2$. Since we are close to the galactic plane the shock wave remains
a plane shock and is given by
\begin{equation}  \label{e.tacc}
   \tacc = \frac{3}{u_1-u_2}
           \left(\frac{\kappa_1}{u_1}+\frac{\kappa_2}{u_2}\right) \: .
\end{equation}
Both diffusion coefficients $\kappa_{1,2} = \kappa(p)$ depend in a Fermi acceleration 
process on
the particle momentum $p$ and can be calculated in the case of 
%%
%% Bohm-Zitat eingefügt
%%
fully developed turbulence from the so-called Bohm-limit \citep{Bo:49}, 
\begin{equation} \label{e.kapp}
   \kappa(p) = \frac{1}{3}\frac{mc^3}{ZeB}
               \frac{p^2}{\sqrt{1+p^2}} \: .
\end{equation}

An energetic particle can be picked up by successive shock waves and become  
accelerated to even higher energies. The typical length scale for the energizing
particle is given by $\kappa/\us$; a particle with different 
energies will experience different compression ratios generated by a changing number of subsequent 
shock waves (cf.~Fig.~\ref{f.shocks}). The spatial size of the shock waves is at least a 
factor of $10^2$ more extended
by changing sizes from a parsec-scale to an almost kiloparsec-scale in galactic winds.
The individual waves merge within a few ${\rm kpc}$
above the galactic plane (see Table~\ref{t.coal}) and hence we expect
only small adiabatic losses compared to acceleration sites located far away in the galactic 
halo, for example~in the outer terminal shock of a galactic wind (\citet{JM:87}).

\begin{figure}   %                                                         Fig. 12
\resizebox{\hsize}{!}{\includegraphics{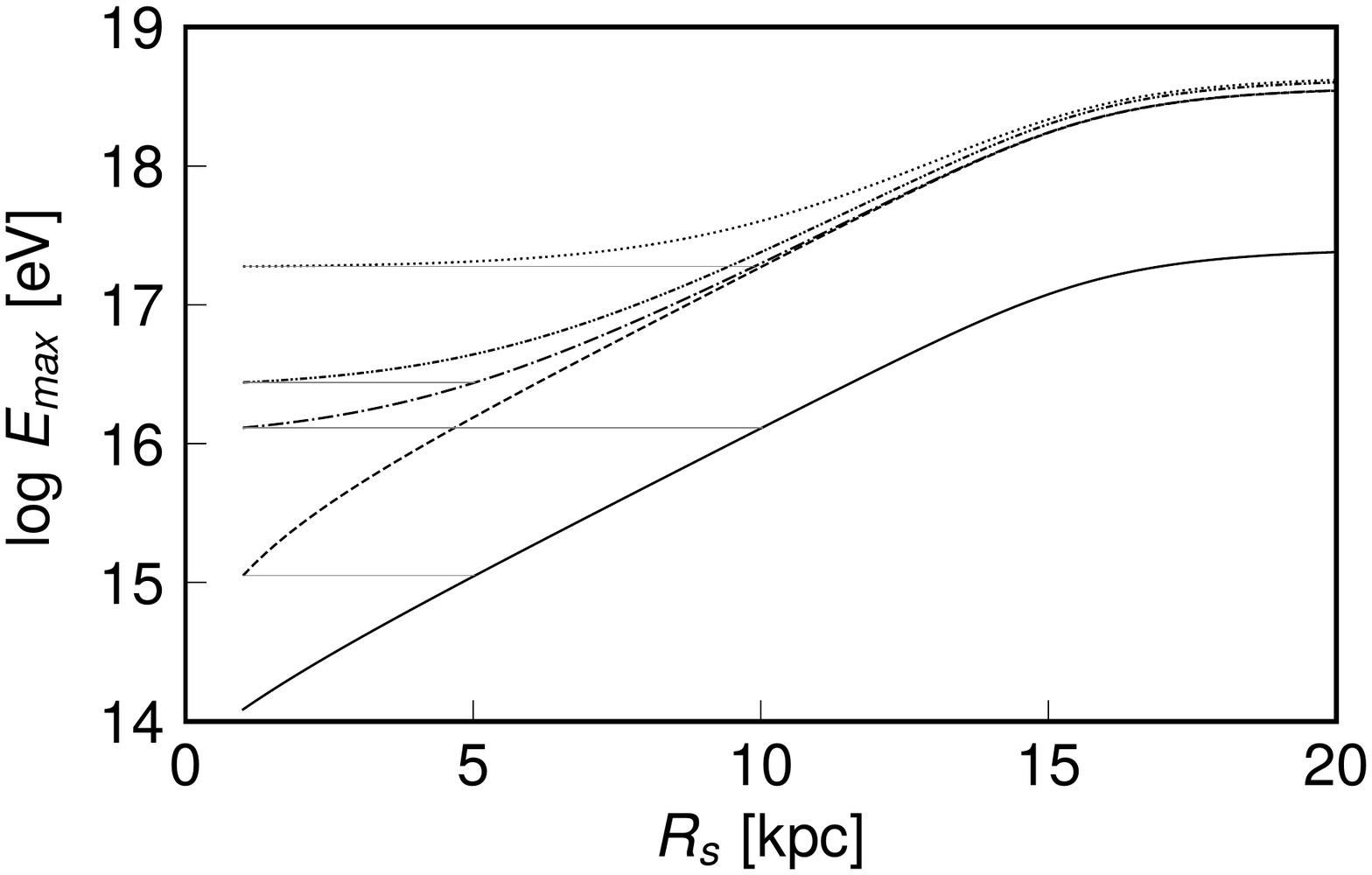}}
  \caption{Evolution of the maximum (full line) particle energy $E_{\rm max}$
           in units of $[{\rm eV}]$ 
           as a function of the shock distance from the galactic plane in units of $[\rm kpc]$
           for {\sf Model H}. The full line corresponds to the first shock 
           (see also Fig.~\ref{f.shocks}), the dashed lines plot particles re-accelerated 
           by the second shock waves, and the dashed-dotted lines exhibit the energy gains
           due to the third shock wave.         }
  \label{f.e_max}
\end{figure}

We emphasize that a
diffusion coefficient on the order of $\kappa\simeq 10^{28}\,{\rm cm^2s^{-1}}$ 
together with a shock speed of $\us\simeq 3000\,{\rm km/s}$
leads to a typical spatial scale of $\kappa/\us\simeq 10\,{\rm pc}$; 
the corresponding acceleration time is 
on the order of $\kappa/\us^2\simeq 3000\,{\rm years}$. If the diffusion
coefficient reaches a maximum value of $\kappa\simeq 10^{30}\,{\rm cm^2s^{-1}}$ 
\citep[see e.g.][]{Axf:81} the spatial scale increases to $1\,{\rm kpc}$ 
with an acceleration time of $3\cdot10^5\,$years. 

At such large particle energies the diffusion
coefficient $\kappa(p)$ increases linearly with the particle momentum $p$
and scales inversely with the local magnetic field. Hence, during the
time before the merging of shock waves $t_{\rm merge}$ (see Table~\ref{t.coal}), 
we expect different particle energies to be exposed to rather different compression ratios and complex flow features.

We concentrate on {\sf Model H,} which mimics an expansion into a static halo.
In Fig. \ref{f.e_max} the maximum particle energy $E_{\rm max}$ is plotted for a particle
starting at the 'knee' with $10^{14}\,{\rm eV}$ (solid line), which gets
accelerated only by the first leading shock wave propagating into a static galactic halo. 
When the last SN explodes after $1.2\cdot 10^7\,$years this leading shock wave has 
travelled a distance of up to $37\,$kpc. The particle energies result from 
integrating Eq.~\ref{e.pmax}.
As can be seen, the main acceleration region is located below $15\,$kpc and 
the first encounter leads to the most important energy increase. Even for 
$z=20\,{\rm kpc}$ the adiabatic losses remain small, since
according to the adopted flux tube geometry we get a geometrical
dilution factor of $1+(20/15)^2 = 2.78$ from Eq.~(\ref{e.az}). The subsequent shock
waves (dashed and dashed-dotted lines) can boost the particle to higher energies but the
low Mach number around $M\simeq 1.2$ limits the acceleration efficiency. The horizontal dotted
lines indicate that we take particles at $5\,$kpc and $10\,$kpc, respectively, to be
re-accelerated by subsequent shocks. Since the efficiency decreases with the
number of individual shocks, we have considered in Fig.~\ref{f.e_max} only the first 
three shock fronts.

As mentioned above, an increase of the magnetic field, that is, $B_0=1\,\mu{\rm G}$ at
the base of our flux tube and/or a different background model
leading to faster shocks, will allow particles to reach energies around the 'ankle'. 
As observations of the non-thermal emission of SNRs suggest
\citep[see e.g.~the case of SN~1006,][]{Ress} the strength of magnetic
fields in the vicinity of a blast wave can significantly be enhanced compared to
typical ISM values. Such amplified magnetic fields could also be present in the
case of shock waves within a galactic wind triggered by repeated SN explosions, 
%%
%% DB: Nebensatz eingefügt; ich denke, dass die magnetische Turbulenz durch die Explosionen erhöht wird.
%%
which, as is known, will generate a high level of magnetic turbulence. 
The physics of such an amplification process together with numerical simulations
is discussed for example~in  \citet{Dow:Dru}. Adopting these higher field strengths,  
energetic particles can be accelerated  to values up to the
'ankle' by a first-order Fermi process in galactic wind shocks. 

The total particle energies will also depend on the background medium and clearly an expansion 
into a halo-like model (cf.~{\sf Model H}) is preferred because an expansion into
an existing galactic wind (cf.~{\sf Model W}) leads to slower but more massive
winds, which can cease shortly after the last SN exploded. An improvement of the calculations presented here, which is the subject of a future paper, will be the effect of the galactic wind and the associated acceleration of CRs on their energy spectrum, which requires their treatment on the distribution function level.

\section{Conclusions}
\label{s.con}
%%
%% DB: hier einige Aspekte, die ich für wichtig halte - bitte prüfen
%%
Our simulations bridge the gap between purely steady-state solutions like in 
\citet{DB91} and time-dependent galactic winds like in \citet{DB1}, by connecting 
outflows, which start at a reference level at $1\, {\rm kpc}$, to the energy 
source of a SB beneath them. Therefore we can quantitatively assess the 
restrictions for steady-state solutions. Since the results we have obtained here show 
that the outflow breaks down on a timescale of a few $10^7$ years after 
switch-off of the SN activity, steady-state flows, which are a simplistic approximation 
to reality, can only exist if the flow time $\tau_{f}$ is less or 
comparable to the SB lifetime plus the break-down time of stationarity. This is given by the distance $d_s$ of the critical point (saddle point singularity in case of outflow) divided by the information carrying signal (sound) speed $c_s$, that is, $\tau_{s} \sim d_c/c_s$. The distance $d_s$ in turn depends on the mass loss rate or the energy input rate, respectively (see \citet{DB91}) and represents the characteristic length scale of influence, because the flow decouples causally beyond the critical point and will no longer be sustained by pressure gradients from below. Within $\tau_s$ the outflow also drains enough material from the SB so that, again, pressure gradients for steady-state flow cannot be maintained for much longer than a flow time. These constraints are somewhat relaxed by the action of centrifugal forces \citep{Zirak:96}, which are not affected by the SB shutdown.

It has been known for some time that low mass loss rates imply high speed winds and vice 
versa (for details see \citet{Bre:94}). High speed winds with low mass loss 
rates will reach the critical point faster, at lower distances from the 
disc, and therefore $\tau_f$ will be smaller and steady-state flows are more 
likely to occur. In contrast massive outflows are sluggish, with low speeds and 
large timescales for acoustic or magnetosonic waves to travel out to the 
critical point. The major difference between thermally- and CR-driven galactic winds is that CRs cannot cool, and they do exhibit a larger scale height,  $|\nabla P_c/P_c|^{-1} >  |\nabla P_g/P_g|^{-1}$ because $\gamma_c < \gamma_g$.
Although each outflow has to be evaluated individually, it seems 
fair to state that SBs with a small stellar content and high gas 
density will have difficulties in developing steady-state flows, provided that 
SB break-out occurs at all. In contrast, SBs with a high 
number of massive stars and a large overpressure with respect to a low density 
ambient medium can develop fast outflows. The most favourable circumstances for 
steady-state flows will be in the case of triggered and/or self-propagating star 
formation, so that channels or cavities that have been created by previous 
generations of SNe can be maintained over a timescale of $\sim 10^8$ years or 
longer. 

The role of the bulk of the CRs in the fluid dynamical approximation, 
coupled by self-excited waves to the plasma, is to provide a softer equation of 
state, that is, $\gamma_c = 4/3$ for a relativistic fluid, thus being less affected by adiabatic losses,  as compared to $\gamma_g = 5/3$ for the thermal plasma. CRs are thereby able to lift off the upper gas layer 
of the halo at larger distances, and accelerate the flow to higher 
velocities, thus promoting steady-state flows with low mass loss rates. 
Interestingly, CR diffusion supports fast winds, as the mass to be 
lifted off is decreased. \citet{FR:18} argue that CR decoupling from the neutral or low ionization gas (due to efficient wave damping in low ionization regions), leads to even faster winds because of the increase of the Alfv\'en speed, particularly near the disc. 
Therefore, depending on the energy-dependent diffusion coefficient, CRs 
will always foster a low-mass- loss-rate, magnetized galactic wind in 
steady-state models as long as MHD scattering waves are present and some coupling to 
the thermal plasma is present.
%
%In general we expect larger outflow velocities with smaller mass loss rates in case of 
%larger diffusion coefficients. 
%This fact results from larger CR pressures at larger
%distances and we emphasize that CRs suffer less from adiabatic losses
%due to $\gc=4/3$ when compared to the thermal gas. 

The advantage of the model, describing CR-driven galactic winds, 
presented here, is that it captures the essential physics of the problem while 
being simple and computationally cheap compared to an ab initio 3D model, 
because it is intrinsically 1D. To keep the flux tube geometry, we had to treat 
the evolution of the SB separately. This raises the question of how to 
connect the SB to the flux tube. Here we have used as a first 
approach constant density and pressure provided over the timescale of 
explosions. In a more realistic treatment, we can calculate these variables 
time dependently. The second simplification, which has to be fixed by hand, is the 
background into which the emanating wind propagates. While this may be the 
subject of a parameter study in a future paper, we have treated here two limiting  
cases, a stationary wind solution and a hydrostatic halo, into which the 
SB plasma and the CRs propagate. 
%
%% % referee suggests rewriting
%
We find that less material located outside the galactic disc (c.f.~{\sf Model H}) leads to a higher asymptotic wind
speed and to a transport of material further outwards.
In essence, these two models should
encompass a large variety of realistic scenarios, and they bear all the features, such as 
complex non-linear wave interactions, which we have discussed here in detail. 
Independent of which of the two limiting models we choose, it is always found that 
CR acceleration predominantly occurs in the region where the flux tubes still maintain planar geometry.
This limits adiabatic losses and leads to an even simpler description. 

%%
%% DB: "ankle" liegt bei 10^18 nicht 10^17; ich denke wir können auch diese Energien erreichen; bitte korrigieren, falls das nicht stimmt
%%
Particle energies up to $10^{18}\, {\rm eV}$, the so-called ankle, can be reached if the flux tube 
contains rather dilute gas close to the static limit, because in this case the Mach number of the shocks will be sufficiently high to boost the energy appreciably, as opposed to a smaller relative velocity and hence Mach number if the background medium is expanding away from the galaxy.
The individual shocks from repeated SN explosions are imprinted on the galactic wind 
and merge at distances less than $5\,{\rm kpc}$ into a single shock wave. This
combined shock wave gets accelerated because of the large density and pressure gradients
and travels at almost constant speeds to significant distances of a few $100\,$kpc. The external pressure is communicated to the outflow, leading to the formation of a reverse shock, which starts to travel inwards until $t_{max} = 9.02\cdot 10^7\,{\rm years}$ and reaches a maximum distance of $z_{max} = 5.1\,{\rm kpc}$.

In summary, it seems plausible that particles can reach energies up to the ankle at $10^{18}\,{\rm eV}$, depending on the diffusion coefficient, which itself is a function of the magnetic field strength. As we have shown, acceleration of energetic particles still occurs sufficiently close to the galactic plane that (i) they do not suffer large adiabatic losses and (ii) they can easily diffuse back to the disc where they are observed. This also explains why CRs below the knee, which are accelerated by disc sources, have a higher degree of isotropy than CRs between the knee and the ankle, which are post-accelerated galactic CRs that are not scattered back from a small diffusion halo of about 1 kpc in z-direction, but from individual local flux tubes, extending several kiloparsecs into the outer halo. The Auger Collaboration \citep{Aug:17} reports the detection of an anisotropy at the $\geq 5.2\sigma$-level for light element particles with energies above 8 $10^{18}$ eV. While an extragalactic origin of these ultra-high-energy CRs is favoured, the actual source(s) still remain(s) a mystery. The CRs post-accelerated in the halo, which we have described here, are typically in the energy range $10^{14} - 10^{18}$ eV, although some contribution to the higher more anisotropic CRs is possible.

\begin{acknowledgements}

This publication is supported by the Austrian Science Fund (FWF).
DB acknowledges partial support from the \emph{Deut\-sche For\-schungs\-ge\-mein\-schaft}, DFG, Priority Programme SPP 1573.
We thank the referee, Joel Bregman, for his encouragement and
suggestions that helped to improve the paper.

\end{acknowledgements}

\bibliographystyle{apj} %% style apj.bst
\bibliography{apj-jour,db1_ref}

\begin{thebibliography}{}
%\bibitem[Adelberger \& Steidel(2000)]{AdSt:00}
%    Adelberger K.~L., Steidel, C.C. 2000, \apj, 544, 218
\bibitem[Amato \& Blasi(2018)]{AB:18} 
Amato, E., Blasi, P., 2018, AdSpR, 62, 2731    
\bibitem[Axford et al.(1977)]{Axf:etal}
    Axford W.I., Leer E., Skadron G. 1977,
    Proc. 15th Int. Cosmic Ray Conf. (Plovdiv) 11, 132
\bibitem[Axford(1981)]{Axf:81}
    Axford W.I. 1981, Proc. Int. School and Workshop on Plasma
    Astrophysics, Varenna, ESA~SP--161, 425
\bibitem[Baumgartner \& Breitschwerdt(2013)]{BB:13}
Baumgartner, V., Breitschwerdt, D., 2013, A\&A, 557,140
\bibitem[Bell(1978a)]{Bel:a}
    Bell A.R. 1978a, \mnras, 182, 147
\bibitem[Bell(1978b)]{Bel:b}
    Bell A.R. 1978b,\mnras, 182, 443
%\bibitem[Bell(1987)]{Bel}
%    Bell A.R. 1987, \mnras, 215, 615
\bibitem[Bell \& Lucek(2001)]{BL:01}
    Bell A.R., Lucek, S.G., 2001, \mnras, 321, 433
%\bibitem[Berezhko(1996)]{Be:96}  
%    Berezhko, E. G. 1996, Astroparticle Phys. 5, 367
%\bibitem[Berezhko et al.(1994)]{Ber:etal}
%    Berezhko E.G., Yelshin V.K., Ksenofontov L.T. 1994,
%    Astroparticle Phys. 2, 215
%\bibitem[Blandford \& Eichler(1987)]{Bla:Eic}
%    Blandford R.D. Eichler, D. 1987, Phys. Rep. 154, 1
\bibitem[Blandford(1988)]{Bla}
    Blandford R.D. 1988, in: Supernova Remnants and the Interstellar
    Medium. Roger R.S., Landecker T.L. (eds.) Cambridge Univ. Press,
    Cambridge, p. 309
\bibitem[Blandford \& Ostriker(1978)]{Bla:Ost:78}
    Blandford R.D., Ostriker, J.P. 1978, \apj, 221, L29
\bibitem[Bohm(1949)]{Bo:49}
   Bohm, D., 1949, The characteristics of electrical discharges in magnetic fields, 1st ed. (New York, McGraw-Hill)
\bibitem[Bregman et al.(1992)]{Bre:etal}
    Bregman J.N., Hogg D.E., Roberts M.S. 1992, \apj, 387, 484    
\bibitem[Bregman(1980)]{Bre:80}
    Bregman J.N., 1980, \apj, 236, 577
\bibitem[Breitschwerdt(2003)]{Br:03} 
    Breitschwerdt D.  2003,  Rev. Mex. Astron. Astrophys., 15, 311    
\bibitem[Breitschwerdt(1994)]{Bre:94} 
    Breitschwerdt D., 1994,  Habilitation Thesis, University of Heidelberg, 158 p.
\bibitem[Breitschwerdt et al.(2002)]{Br:02}
    Breitschwerdt D., Dogiel, V.A., V{\" o}lk, H.J. 2002, \aap, 385, 216
\bibitem[Breitschwerdt et al.(2016)]{BF:16}
    Breitschwerdt D., Feige, J., Schulreich, M.M., de Avillez, M.A., Dettbarn, 
C., Fuchs, B. 2016, \nat, 532, 73
\bibitem[Breitschwerdt et al.(1991)]{DB91}
    Breitschwerdt D., McKenzie J.F., V\"olk H.J. 1991, A\&A, 245, 79 
\bibitem[Breitschwerdt et al.(1993)]{DB93}
    Breitschwerdt D., McKenzie J.F., V\"olk H.J. 1993, A\&A, 269, 54
\bibitem[Breitschwerdt \& Schmutzler(1994)]{BS:94}
    Breitschwerdt D., Schmutzler M. 1994, \nat, 371, 774
\bibitem[Breitschwerdt \& Schmutzler(1999)]{BS:99}
    Breitschwerdt D., Schmutzler M. 1999, A\&A, 347, 650
\bibitem[Canizares et al.(1987)]{Can:etal}
    Canizares C.R., Fabbiano G., Trinchieri G. 1987, \apj, 312, 503
\bibitem[Courant \& Friedrichs(1948)]{CF}
    Courant R., Friedrichs K.O. 1948, Supersonic Flow and Shock Waves, Springer-Verlag
    New York, 1948    
\bibitem[Dawson et al.(2002)]{Daw:02}
    Dawson S., Spinrad H., Stern D., Dey A., van Breugel W., de Vries W., Reuland M.
    2002, \apj, 570, 92
\bibitem[Avillez(2000)]{deA:00}
    de Avillez M. 2000, \apss~272, 22
\bibitem[Avillez \& Breitschwerdt(2004)]{AB:04}
    de Avillez M., Breitschwerdt D., 2004, A\&A, 425, 899  
\bibitem[Avillez \& Breitschwerdt(2005)]{AB:05}
    de Avillez M., Breitschwerdt D., 2005, A\&A, 436, 585  
\bibitem[Avillez \& Breitschwerdt(2005)]{AB:07}
    de Avillez M., Breitschwerdt D., 2007, \apjl, 665, 35  
\bibitem[Dogiel et al.(1994)]{Dog:etal:94}
    Dogiel, V.A., Gurevich, A.V., Zybin, K.P., 1994, A\&A, 281, 937
%\bibitem[Dorfi(1990)]{Dor:90}
%    Dorfi E.A. 1990, A\&A, 234, 419
%\bibitem[Dorfi(1991)]{Dor:91}
%    Dorfi E.A. 1991, A\&A, 251, 597
%\bibitem[Dorfi(1998)]{Saas}
%    Dorfi E.A. 1998, in: Computational Methods for Astrophysical Fluid Flows,
%    27th Saas Fee Course, Springer, Berlin, p. 263
\bibitem[Dorfi \& Breitschwerdt(2012)]{DB1}
    Dorfi, E.A., Breitschwerdt D., 2012, A\&A, 540, A77 
\bibitem[Dorfi \& Drury(1987)]{DD}
    Dorfi E.A., Drury L.O'C. 1987, J. Comp. Phys. 69, 175 
\bibitem[Downes \& Drury(2014)]{Dow:Dru}
    Downes T.P., Drury L.O'C. 2014, \mnras, 444 365
\bibitem[Drury(1983)]{Dru}
    Drury L.O'C. 1983, Rep. Prog. Phys. 46, 973   
\bibitem[Drury \& V\"olk(1981)]{Dru:Voe}
    Drury L.O'C., V\"olk H.J. 1981, \apj, 248, 344
%\bibitem[Drury et al.(2003)]{Dru:etal:03}
%    Drury L.O'C, van der Swaluw, E., Carroll, O., 2003, astro-ph/0309820
%\bibitem[Dubois \& Teyssier(2010)]{DT:10}
%   Dubois, Y., Teyssier, R., 2010, A\&A, 523, 72
%\bibitem[Eriksen et al.(2011)]{Eriksen}
%    Eriksen A.K.,  Hughes J.P., Badenes C., Fesen R., Ghavamian P., 
%    Moffett D., Plucinksy P.P., Rakowski C.E., Reynoso E.M., Slane P. 2011, \apjl, 728, 28
%\bibitem[Everett et al.(2010)]{Ev:10}
%    Everett, J.E., Schiller, Q.G., Zweibel, E.G. 2010, ApJ, 711, 13 
%\bibitem[Everett et al.(2008)]{Ev:08}
%    Everett, J.E., Zweibel, E.G., Benjamin, R.A. et al. 2008, \apj, 674, 285  
%\bibitem[Fichtner etal.(1990)]{Ficht:etal:90}
%   Fichtner H., Fahr H.J., Neutsch W., Schlickeiser R., Crusius-W\"atzel, A., 
%   Lesch H. 1990, Nuovo Cimento B, 106(8), 909
%\bibitem[Fichtner et al.(1991)]{Ficht:91}
%    Fichtner H., Neutsch W., Fahr H.J., Schlickeiser R. 1991, \apj, 371, 98
%\bibitem[Franco et al.(1991)]{Franco:91}
%    Franco J., Ferrini F., Ferrara A., Barsella B. 1991 \apj 366 443   
\bibitem[Dyson \& Williams(1997)]{DW:97}
Dyson, J.E., Williams, 1997, The Physics of the Interstellar Medium, ch. 7.1.10, Institute of Physics Publishing
\bibitem[Farber et al.(2018)]{FR:18}
Farber, R., Ruszkowski, M., Yang, H.-Y. K., Zweibel, E. G. 2018, \apj, 856, 112
\bibitem[Fuchs et al.(2006)]{FB:06}
    Fuchs, B., Breitschwerdt, D., de Avillez, M.A., Dettbarn, C., Flynn, C., 
2006, \mnras 373, 993
\bibitem[Ginzburg \& Ptuskin(1985)]{Gin:Pu}
    Ginzburg V.L., Ptuskin V.S. 1985, Sov.Sci.Rev.E. \apss, 4, 161
\bibitem[Girichidis et al.(2018)]{GN:18}
Girichidis, P., Naab, T., Hanasz, M., Walch, S., 2018, \mnras, 479, 3042
\bibitem[Gressel et al.(2008)]{Gr:etal:08}
    Gressel, O., Elstner, D., Ziegler, U. R\"udiger, G., A\&A, 486, L35
\bibitem[Habe \& Ikeuchi(1980)]{HI:80}
    Habe A., Ikeuchi S. 1980, Rep.Prog.Theor.Phys. 64, 1995
\bibitem[Hanasch et al.(2013)]{Han:etal}
   Hanasz M., Lesch H.,Naab T., Gawryszczak A., Kowalik K., W\'olt\'anski D. 2013, \apj 777, L38
\bibitem[Hartquist(1983)]{Ha:83}
    Hartquist T. 1983, \mnras, 203, 117
\bibitem[Heesen et al.(2018)]{HK:18}
Heesen, V., Krause, M., Beck, R., Adebahr, B,; Bomans, D. J., Carretti, E., Dumke, M., Heald, G., Irwin, J., Koribalski, B. S., Mulcahy, D. D., Westmeier, T., Dettmar, R.-J. 2018, MNRAS 476, 158
\bibitem[Ipavich(1975)]{Ip:75}
    Ipavich F. 1975, \apj, 196, 107 
\bibitem[Jokipii(1987)]{Jo:87}
    Jokipii J.R. 1987, \apj, 313, 842  
\bibitem[Jokipii \& Morfill(1985)]{JM:85}
    Jokipii J.R., Morfill G.E. 1985, \apj, 290, L1    
\bibitem[Jokipii \& Morfill(1987)]{JM:87}
    Jokipii J.R., Morfill G.E. 1987, \apj, 312, 170    
\bibitem[Kafatos \& McCray(1987)]{KM:87}
    Kafatos, M., McCray, R. 1987, ApJ 317, 190
\bibitem[Kahn(1981)]{Kahn}
    Kahn F.D. 1981, in: Investigating the Universe, F.D. Kahn (ed.),
    D. Reidel Publ. Comp., Dordrecht, p. 1
\bibitem[Kapferer et al.(2006)]{Kap:06}
    Kapferer W., Ferrari, C., Domainko, W., Mair, M., Kronberger, T., Schindler, S., Kimenswenger, S., 
    van Kampen, E., Breitschwerdt, D., Ruffert, M. 2006, A\&A, 447, 827
\bibitem[Kikuchi et al.(1999)]{Kik:99}
    Kikuchi K., Furusho T., Ezawa H., Yamasaki N.Y., Ohashi T., Fukazawa Y., Ikebe Y. 1999, 
    \pasj, 51, 301
\bibitem[Ko(1991)]{Ko}
    Ko C.M. 1991, A\&A, 242, 85 
\bibitem[Kompaneets(1960)]{Kompa}
    Kompaneets A.S. 1960, Soviet Phys. Doklady, 5, 46
\bibitem[Korpi et al.(1999)]{Ko:etal:99}
    Korpi, M.J., Brandenburg, A., Shukurov, A., Tuominen, I., Nordlund, A., \apjl, 514, 99 
\bibitem[Krymsky(1977)]{Kry}
    Krymsky G.F. 1977, Dokl.~Nauk.~SSR~234, 1306,
    (Engl.~trans. Sov.~Phys.~Dokl.~23, 327)
\bibitem[Kulpa-Dybel et al.(2011)]{Kulpa}
    Kulpa-Dybell K., Otmianowska-Mazur K., Kulesza-{\.Z}ydzik B., Hanasz M., Kowal G.,
    W{\'o}lta{\'n}ski D., Kowalik K. 2011, \apj, 733, L18
\bibitem[Kulsrud \& Pearce(1969)]{Kul:Pea}
    Kulsrud R.M., Pearce W.D. 1969, \apj, 156, 445
\bibitem[Lagage \& Cesarsky(1983)]{LC:83}
    Lagage P.O, Cesarsky C.J. 1983, A\&A, 125, 249L
%\bibitem[Lerche \& Schlickeiser(1982)]{LS:82}
%    Lerche I., Schlickeiser R. 1982, \aap, 107, 148
\bibitem[LeVeque(1990)]{LeV}
    LeVeque, R.J. 1990, Numerical Methods for Conservation Laws,  
    Birkh\"auser-Verlag, Basel
%\bibitem[Loewenstein(2001)]{Loew:01}
%    Loewenstein, M. 2001, \apj, 557, 573
%\bibitem[Lucek \& Bell(2000)]{LB:00}    
%    Lucek S.G., Bell A.R. 2000, \mnras, 314, 65L
\bibitem[Lochhaas et al.(2018)]{LT:18}
Lochhaas, C., Thompson, T.A., Quataert, E., Weinberg, D.H., 2018, \mnras, 481, 1873
\bibitem[Maciejewski \& Cox(1999)]{Mac}
    Maciejewski W., Cox D.P. 1999, \apj, 792
%\bibitem[Markiewicz et al.(1990)]{Mar:etal}
%    Markiewicz W.J., Drury L.O'C, V\"olk H.J. 1990, A\&A, 236, 487
\bibitem[Mao \& Ostriker(2018)]{MO:18}
Mao, S. A., Ostriker, E.C., 2018, \apj, 854,89
\bibitem[McKee(1988)]{McK}
    McKee C.F. 1988, in: Supernova Remnants and the Interstellar
    Medium. Roger R.S., Landecker T.L. (eds.) Cambridge Univ. Press,
    Cambridge, p. 205 
\bibitem[McKee \& Cowie(1977)]{McK:Cow}
    McKee C.F., Cowie L. 1977, \apj, 215, 213
\bibitem[McKee \& Ostriker(1977)]{McK:Ost}
    McKee C.F., Ostriker J.P. 1977, \apj, 218, 148
\bibitem[McKenzie \& V\"olk(1982)]{MV:82}
    McKenzie, J.F., V\"olk, H.J. 1982, A\&A, 116, 191
\bibitem[Miyamoto \& Nagai(1975)]{MN}
    Miyamoto M., Nagai R. 1975, PASJ 27, 533
\bibitem[Molendi et al.(1999)]{Mol:99}
    Molendi S., de Grandi S., Fusco-Femiano R., Colafrancesco S., Fiore F., Nesci R.,
    Tamburelli F. 1999, \apjl, 525, L73    
\bibitem[Normandeau et al.(1996)]{Nor:96}
    Normandeau M., Taylor A.R., Dewdney P.E. 1996, \nat, 380, 687 
\bibitem[Pais et al.(2018)]{PP:18}
    Pais, M., Pfrommer, C., Ehlert, K., Pakmor, R. 2018, \mnras 478, 5278
\bibitem[Parker(1992)]{Pa:92}
   Parker, E.N., \apj, 401, 137
\bibitem[Parker(1958)]{Par:58}
    Parker, E.N. 1958, \apj, 128, 664 
\bibitem[Pierre Auger Collaboration et al.(2017)]{Aug:17}
   P. A. Collaboration et al., 2017, Science, 357, 1266
\bibitem[Ponman et al.(1999)]{Pon:99}
    Ponman T.J., Cannon D.B., Navarro J.F. 1999, \nat, 397, 135
\bibitem[Ptuskin(1997)]{Ptus:97}
    Ptuskin V.S. 1997, Adv. in Space Res. 19, 697
\bibitem[Ptuskin et al.(1997)]{Ptus:etal:97} 
    Ptuskin V.S., V\"olk, H.J., Zirakashvili, V.N., Breitschwerdt, D., A\&A, 321, 434  
\bibitem[Ressler et al.(2014)]{Ress}
    Ressler, S.M., Katsuda, S-, Reynolds, S.P., Long. K.S., Petre, R., 
    Williams, B.J., Winkler, P.F., \apj 790, 85   
\bibitem[Rupke(2018)]{ru:18}
   Rupke, D. 2018, Galaxies 6, 138
\bibitem[Sharma et al.(2017)]{sharma+17}  
     Sharma, M., Theuns, T., Frenk, C., Bower, R.G., Crain, R.A., Schaller, M., 
     Schaye, J., 2017, \mnras, 468, 2176
\bibitem[Ruszkowski et al.(2017)] {Rus:2017}
    Ruszkowski, M., Yang ,H.-Y. K.,  Zweibel, E., 2017, \apj 834, 208  
%\bibitem[Reynolds et al.(2001)]{Rey:01}
%    Reynolds R.J., Sterling N.C., Haffner L.M., 2001, \apjl, 558, L101
%\bibitem[Sarazin \& White(1988)]{Sar:Whi}
%    Sarazin C.L., White R.E. 1988, \apj, 331, 102
%\bibitem[Soida et al.(2011)]{Soida}
%    Soiada M., Krause M., Dettmar R.-J., Urbanik M. 2011, A\&A, 531, 127
%\bibitem[Tammann(1982)]{Tam}
%    Tammann G. 1982, in: Supernova: A survey of current research,
%    Eds. M. Ress and R. Stoneham, Dordrecht, Reidel, p. 371
%\bibitem[Tscharnuter \& Winkler(1979)]{TW}
%    Tscharnuter W.M., Winkler K.-H.A. 1979, 
%    Comp.Phys.Comm. 19, 171
%\bibitem[V\"olk(1987)]{Voe:87}
%    V\"olk H.J. 1987,
%   Proc.~20th Int.~Cosmic Ray Conf.~(Moscow) 7, 157
\bibitem[Skinner \& Ostriker(2015)]{SO:15}
Skinner, M.A., Ostriker, E.C. 2015, \apj, 809, 187
\bibitem[Thomas \& Pfrommer(2019)]{TP:19}
Thomas, Pfrommer, C., 2019, \mnras, 485, 2977
\bibitem[V\"olk \& Biermann (1988)]{VB:88}
    V\"olk, H.J., Biermann, P.L., 1988, \apj, 333, 65
\bibitem[V\"olk et al.(1984)]{Voe:etal}
    V\"olk H.J., Drury L.O'C., McKenzie J.F. 1984, A\&A, 130, 19
%\bibitem[V\"olk \& McKenzie(1981)]{Voe:mck}
%    V\"olk H.J., McKenzie J.F. 1981,
%    Proc.~17th Int.~Cosmic Ray Conf.~(Paris) 9, 246
\bibitem[V\"olk \& Zirakashvili(2004)]{VZ:04}
    V\"olk H.J, Zirakashvili, V.N. 2004, A\&A, 417, 807
\bibitem[Zirakashvili et al.(1996)]{Zirak:96}
   Zirakashvili, V.N., Breitschwerdt, D., Ptuskin V.S., V\"olk, H.J. 1996, A\&A, 311, 113  
\bibitem[Zweibel(2017)]{ZW:17}
    Zweibel, E.G. 2017, Phys. of Plasmas, 24, 055402
%

\end{thebibliography}

\end{document}